\documentclass{article}
\usepackage{jcappub}
\usepackage{amsmath,graphicx,capt-of,ifthen,calc}
\usepackage[numbers,sort&compress]{natbib}
\usepackage{setspace,comment,bm,subfig,graphics}
\usepackage{caption}
\usepackage{color}
\usepackage{url}

\usepackage{mciteplus,enumerate}
\usepackage{revsymb}
\usepackage{dcolumn}

\newcommand{\tmop}[1]{\ensuremath{\operatorname{#1}}}

\newcommand{\lame}{\lambdabar_e}

\newcommand{\dd}{{\rm d}}

\newcommand{\gravitons}{KK gravitons}
\newcommand{\frachoriz}[2]{{#1}\mathord{\left/{\vphantom {{#1}{#2}}} \right. \kern-\nulldelimiterspace} {#2}}
\newcolumntype{d}[1]{D{.}{.}{#1}} 

\title{Limits on Large Extra Dimensions Based on Observations of Neutron Stars with the \emph{Fermi}-LAT}
\collaboration{Fermi-LAT Collaboration}%\\Author list created Sunday 28 Aug 2011 17:05 PDT}
\author[a]{M. Ajello}\affiliation[a]{Stanford University, Applied Physics Department, Physics Department, SLAC National Accelerator Laboratory, W. W. Hansen Experimental Physics Laboratory}
\author[b]{L. Baldini}\affiliation[b]{INFN and University of Pisa}
\author[c]{G. Barbiellini}\affiliation[c]{INFN Trieste}
\author[d]{D. Bastieri}\affiliation[d]{INFN Padova}
\author[a]{K. Bechtol}%\affiliation[a]{Stanford University, Physics Department}
\author[b]{R. Bellazzini}%\affiliation{INFN and University of Pisa}
\author[a]{B. Berenji}\emailAdd{bijanb@alumni.stanford.edu}%\affiliation[g]{Stanford University, Applied Physics Department}
\author[a]{E.D. Bloom}\emailAdd{elliott@slac.stanford.edu}%\affiliation{SLAC National Accelerator Laboratory, Stanford University}
\author[e]{E. Bonamente}\affiliation[e]{INFN Perugia}
\author[a]{A.W. Borgland}%\affiliation{SLAC National Accelerator Laboratory, Stanford University}
\author[b]{J. Bregeon}%\affiliation{INFN and University of Pisa}
\author[f]{M. Brigida}\affiliation[f]{Dipartimento Interateneo di Fisica dell'Universit\`a e Politecnico di Bari and INFN Sezione di Bari}
\author[g]{P. Bruel}\affiliation[g]{Laboratoire Leprince-Ringuet, \'Ecole polytechnique}
\author[a]{R. Buehler}%\affiliation{SLAC National Accelerator Laboratory, Stanford University}
\author[d]{S. Buson}%INFN Padova 
\author[h]{G.A. Caliandro}\affiliation[h]{Institut de Ci\`encies de l'Espai}
\author[a]{R.A. Cameron}%\affiliation{SLAC National Accelerator Laboratory, Stanford University}
\author[i]{P.A. Caraveo}\affiliation[i]{INAF-IASF, Milano}
\author[j]{J.M. Casandjian}\affiliation[j]{Laboratoire AIM, Saclay}
\author[e]{C. Cecchi}%\affiliation[m]{INFN Perugia}
\author[a]{E. Charles}%\affiliation{SLAC National Accelerator Laboratory, Stanford University}
\author[k]{A. Chekhtman}\affiliation[k]{Artep, Inc.}
\author[a]{J. Chiang}%\affiliation{SLAC National Accelerator Laboratory, Stanford University}
\author[l]{S. Ciprini}\affiliation[l]{ASI Science Data Center}
\author[a]{R. Claus}%\affiliation[a]{SLAC National Accelerator Laboratory, Stanford University}
\author[m]{J. Cohen-Tanugi}\emailAdd{johann.cohen-tanugi@lupm.in2p3.fr}\affiliation[m]{Laboratoire Univers et Particules de Montpellier}
\author[n]{J. Conrad}\affiliation[n]{Stockholms Universitet}
\author[l]{S. Cutini}%\affiliation[o]{Agenzia Spaziale Italiana Science Data Center}
\author[o]{A. de Angelis}\affiliation[o]{INFN and University of Udine}
\author[f]{F. de Palma}%\affiliation{Dipartimento Interateneo di Fisica dell'Universit\`a e Politecnico di Bari and INFN Sezione di Bari}
\author[p]{C.D. Dermer}\affiliation[p]{Naval Research Laboratory, Space Science Division}
\author[a]{E. do Couto e Silva}%\affiliation{SLAC National Accelerator Laboratory, Stanford University}
\author[a]{P.S. Drell}%\affiliation{SLAC National Accelerator Laboratory, Stanford University}
\author[a]{A. Drlica-Wagner}%\affiliation[a]{Stanford University, Physics Department}
\author[a]{T. Enoto}%\affiliation{SLAC National Accelerator Laboratory, Stanford University}
\author[f]{C. Favuzzi}%\affiliation{Dipartimento Interateneo di Fisica dell'Universit\`a e Politecnico di Bari and INFN Sezione di Bari}
\author[g]{S.J. Fegan}%\affiliation{Laboratoire Leprince-Ringuet, \'Ecole polytechnique}
\author[q]{E.C. Ferrara}\affiliation[q]{NASA Goddard Space Flight Center}
\author[r]{Y. Fukazawa}\affiliation[r]{Hiroshima University, Physical Sciences Department}
\author[f]{P. Fusco}%\affiliation{Dipartimento Interateneo di Fisica dell'Universit\`a e Politecnico di Bari and INFN Sezione di Bari}
\author[s]{F. Gargano}\affiliation[s]{INFN Sezione di Bari}
\author[l]{D. Gasparrini}%\affiliation{Agenzia Spaziale Italiana Science Data Center}
\author[e]{S. Germani}%\affiliation{INFN Perugia}
\author[f]{N. Giglietto}%\affiliation{Dipartimento Interateneo di Fisica dell'Universit\`a e Politecnico di Bari and INFN Sezione di Bari}
\author[f]{F. Giordano}%\affiliation{Dipartimento Interateneo di Fisica dell'Universit\`a e Politecnico di Bari and INFN Sezione di Bari}
\author[t]{M. Giroletti}\affiliation[t]{INAF Istituto di Radioastronomia, Bologna}
\author[a]{T. Glanzman}%\affiliation[a]{SLAC National Accelerator Laboratory, Stanford University}
\author[a]{G. Godfrey}%%%%%%%%%%\affiliation[a]{SLAC National Accelerator Laboratory, Stanford University}
\author[a]{P. Graham}%\affiliation{Stanford University, Physics Department}
\author[l]{I.A. Grenier}%\affiliation{Laboratoire AIM, Saclay}
\author[u]{S. Guiriec}\affiliation[u]{University of Alabama in Huntsville}
\author[d]{M. Gustafsson}%\affiliation{INFN Padova}
\author[k]{D. Hadasch}%\affiliation{Institut de Ci\`encies de l'Espai}
\author[a]{M. Hayashida}%\affiliation{SLAC National Accelerator Laboratory, Stanford University}
\author[v]{R.E. Hughes}\affiliation[v]{The Ohio State University}
\author[a]{A.S. Johnson}%\affiliation{SLAC National Accelerator Laboratory, Stanford University}
\author[a]{T. Kamae}%\affiliation{SLAC National Accelerator Laboratory, Stanford University}
\author[w]{H. Katagiri}\affiliation[w]{Ibaraki University}
\author[x]{J. Kataoka}\affiliation[x]{Waseda University}
\author[y]{J. Kn\"odlseder}\affiliation[y]{IRAP CNRS}
\author[b]{M. Kuss}%\affilation[b]{INFN and University of Pisa}
\author[a]{J. Lande}%\affiliation{Stanford University, Physics Department}
\author[b]{L. Latronico}%\affiliation{INFN and University of Pisa}
\author[z]{A.M. Lionetto}\affiliation[z]{INFN Roma Tor Vergata}
\author[c]{F. Longo}%\affiliation[c]{INFN Trieste}
\author[f]{F. Loparco}%\affiliation{Dipartimento Interateneo di Fisica dell'Universit\`a e Politecnico disBari and INFN Sezione di Bari}
\author[p]{M.N. Lovellette}%i\affiliation{Naval Research Laboratory, Space Science Division}
\author[e]{P. Lubrano}\affiliation[h]{INFN Perugia}
\author[s]{M.N. Mazziotta}%\affiliation[s]{INFN Sezione di Bari}
\author[a]{P.F. Michelson}%\affiliation[a]{Stanford University, W. W. Hansen Experimental Physics Laboratory}
\author[a]{W. Mitthumsiri}%\affiliation{Stanford University, Physics Department}
\author[r]{T. Mizuno}%\affiliation{Hiroshima University, Physical Sciences Department}
\author[f]{C. Monte}%\affiliation{Dipartimento Interateneo di Fisica dell'Universit\`a e Politecnico di Bari and INFN Sezione di Bari}
\author[a]{M.E. Monzani}%\affiliation{SLAC National Accelerator Laboratory, Stanford University}
\author[z]{A. Morselli}%\affiliation{INFN Roma Tor Vergata}
\author[a]{I.V. Moskalenko}%\affiliation{Stanford University, W. W. Hansen Experimental Physics Laboratory}
\author[a]{S. Murgia}%\affiliation{SLAC National Accelerator Laboratory, Stanford University}
\author[z]{J.P. Norris}\affiliation[z]{Boise State University}
\author[m]{E. Nuss}%\affiliation{Laboratoire Univers et Particules de Montpellier}
\author[r]{T. Ohsugi}%\affiliation{Hiroshima University, Astrophysical Science Center}
\author[a]{A. Okumura}%\affiliation{SLAC National Accelerator Laboratory, Stanford University}
\author[a]{E. Orlando}%\affiliation{Stanford University, W. W. Hansen Experimental Physics Laboratory}
\author[aa]{J.F. Ormes}\affiliation[aa]{University of Denver}
\author[ab]{M. Ozaki}\affiliation[ab]{Institute of Space and Aeronautical Science (JAXA)}
\author[ac]{D. Paneque}\affiliation[ac]{Max-Planck-Institut f\"ur Physik, M\"unchen}
\author[b]{M. Pesce-Rollins}%\affiliation{INFN and University of Pisa}
\author[m]{M. Pierbattista}%\affiliation{Laboratoire AIM, Saclay}
\author[m]{F. Piron}%\affiliation{Laboratoire Univers et Particules de Montpellier}
\author[ad]{G. Pivato}\affiliation[ad]{University of Padova, Fisica}
\author[f]{S. Rain\`o}%\affiliation{Dipartimento Interateneo di Fisica dell'Universit\`a e Politecnico di Bari and INFN Sezione di Bari}
\author[b]{M. Razzano}%\affiliation{INFN and University of Pisa}
\author[ae]{S. Ritz}\affiliation[ae]{University of California, Santa Cruz Institute for Particle Physics}
\author[af]{M. Roth}\affiliation[af]{University of Washington}
\author[ae]{P.M. Saz Parkinson}%\affiliation{University of California, Santa Cruz Institute for Particle Physics}
\author[ah]{J.D. Scargle}\affiliation[ah]{NASA Ames Research Center}
\author[ai]{T.L. Schalk}%\affiliation[ai]{University of California, Santa Cruz Institute for Particle Physics}
\author[b]{C. Sgr\`o}\affiliation[b]{INFN and University of Pisa}
\author[aj]{E.J. Siskind}\affiliation[aj]{NYCB Real-Time Computing Inc.}
\author[b]{G. Spandre}%\affiliation{INFN and University of Pisa}
\author[f]{P. Spinelli}%\affiliation{Dipartimento Interateneo di Fisica dell'Universit\`a e Politecnico di Bari and INFN Sezione di Bari}
\author[ak]{D.J. Suson}\affiliation[ak]{Purdue University - Calumet}
\author[a]{H. Tajima}%\affiliation{SLAC National Accelerator Laboratory, Stanford University}
\author[al]{H. Takahashi}\affiliation[al]{Hiroshima University, Astrophysical Science Center}
\author[a]{T. Tanaka}%\affiliation{SLAC National Accelerator Laboratory, Stanford University}
\author[a]{J.G. Thayer}%\affiliation{SLAC National Accelerator Laboratory, Stanford University}
\author[a]{J.B. Thayer}%\affiliation{SLAC National Accelerator Laboratory, Stanford University}
\author[d]{L. Tibaldo}%\affiliation{INFN Padova}
\author[b]{M. Tinivella}%\affiliation{INFN and University of Pisa}
\author[h]{D.F. Torres}%\affiliation{Institut de Ci\`encies de l'Espai}
\author[q]{E. Troja}%\affiliation[q]{NASA Goddard Space Flight Center}
\author[a]{Y. Uchiyama}%\affiliation[a]{SLAC National Accelerator Laboratory, Stanford University}
\author[a]{T.L. Usher}%\affiliation{SLAC National Accelerator Laboratory, Stanford University}
\author[a]{J. Vandenbroucke}%\affiliation{SLAC National Accelerator Laboratory, Stanford University}
\author[m]{V. Vasileiou}%\affiliation{Laboratoire Univers et Particules de Montpellier}
\author[a]{G. Vianello}%\affiliation{SLAC National Accelerator Laboratory, Stanford University}
\author[z]{V. Vitale}%\affiliation{INFN Roma Tor Vergata}
\author[a]{A.P. Waite}%\affiliation{SLAC National Accelerator Laboratory, Stanford University}
\author[v]{B.L. Winer}%\affiliation{The Ohio State University}
\author[p]{K.S. Wood}%\affiliation{Naval Research Laboratory, Space Science Division}
\author[a]{M. Wood}%\affiliation{SLAC `National Accelerator Laboratory, Stanford University}
\author[n]{Z. Yang}%\affiliation{Stockholms Universitet}
\author[n]{S. Zimmer}%\affiliation{Stockholms Universitet}

\abstract{We present limits for the compactification scale in the theory of Large Extra Dimensions (LED) proposed by Arkani-Hamed, Dimopoulos, and Dvali.  We use 11 months of data from the Fermi Large Area Telescope (Fermi-LAT) to set gamma ray flux limits for 6 gamma-ray faint neutron stars (NS). To set limits on LED we use the model of Hannestad and Raffelt (HR) that calculates the Kaluza-Klein (KK) graviton production in supernova cores and the large fraction subsequently gravitationally bound around the resulting NS. The predicted decay of the bound KK gravitons to $\gamma\gamma$ should contribute to the flux from NSs. Considering 2 to 7 extra dimensions of the same size in the context of the HR model, we use Monte Carlo techniques to calculate the expected differential flux of gamma-rays arising from these KK gravitons, including the effects of the age of the NS, graviton orbit, and absorption of gamma-rays in the magnetosphere of the NS. We compare our Monte Carlo-based differential flux to the experimental differential flux using maximum likelihood techniques to obtain our limits on LED. Our limits are more restrictive than past EGRET-based optimistic limits that do not include these important corrections. Additionally, our limits are more stringent than LHC based limits for 3 or fewer LED, and comparable for 4 LED. We conclude that if the effective Planck scale is around a TeV, then for 2 or 3 LED the compactification topology must be more complicated than a torus.}
\begin{document}
%\graphicspath{finalGraphics}
\date{\today}
\maketitle
%\End{comment}
%\include{tempabs}                                                                                   
%\maketitle

\section{Introduction}
	In the Standard Model of particle physics, gravity is not unified with the other 3 fundamental forces .This is manifested by the \emph{hierarchy problem}, the fact that the electroweak mass scale $M_{\tmop{EW}} \sim 1 \tmop{TeV}$ is many orders of magnitude smaller than the Planck mass scale $M_{\tmop{P}} \approx 1.22 \times 10^{16} \tmop{TeV}$ {\cite{ADD98}}.   Arkani-Hamed, Dimopoulos, and Dvali (ADD) propose a model of Large Extra Dimensions (LED) as a solution for the hierarchy problem.  The ADD scenario may be embedded into string theory, which allows for the existence of compactified extra dimensions.  Due to the presence of $n$ extra dimensions, at length scales smaller than the size of the extra dimensions, the gravitational potential between test masses has a $1 / r^{n+1}$ dependence; however, on scales larger than the size of the extra dimensions the gravitational potential reverts to the ordinary $1 / r$ dependence. For a given $n$, if all the extra dimensions are toroidally compactified, \emph{i.e.}, have the same size $R$, the effective Planck mass in the $(n + 4)-$dimensional space, $M_D$, is related to the reduced Planck mass $\bar{M}_P = M_P/\sqrt{8\pi}$ by the relation: 

\begin{equation} \bar{M}^2_P = R^n M^{n+2}_{D}. \end{equation}

In the ADD model, the hierarchy problem is solved because the presence of LED  brings the effective Planck mass, $M_D$, to the TeV scale, the truly fundamental scale of gravity.  As a consequence, the associated Kaluza-Klein gravitons, denoted by $G_{KK}$, are massless in the bulk, but they have mass on the 3-brane related to their momentum in the bulk (unlike the gravitons of General Relativity).%, which is related to the $G_{KK}$ momentum in the additional extra dimensions\cite{PDG2010}.

According to the ADD model, it is possible to place constraints on extra dimensions by $G_{KK}$ emission from nucleon-nucleon gravibremsstrahlung in type II supernova cores, 
$N N \rightarrow N N G_{KK}$. \ ADD obtain limits from Supernova (SN) 1987A, assuming pion-exchange mediated by the strong force as the dominant 
process. Hanhart, Reddy and Savage (2001) assume a different process\cite{hanhart-2001}.  They use nucleon-nucleon gravibremsstrahlung  
mediated by nucleon-nucleon scattering to obtain the emission rate for KK gravitons.  Furthermore, they indicate that the actual details of the scattering 
process are not important in the soft-radiation limit, where the energy of the outgoing gravitons is much less than the energy other incoming nucleons{\cite{hanhart-2001}}. Then they proceed to obtain limits for $n = 2$ and $n = 3$ extra dimensions from SN1987A.  This is based on the argument that the observed neutrino luminosity sets an upper bound of  $10^{19}$ ergs g$^{-1}$s$^{-1}$ on the energy loss rate into particles other than neutrinos such as $G_{KK}$\cite{raffeltBook}.

Hannestad and Raffelt (henceforth HR \cite{HR2003}) extend this idea to neutron stars, proposing that if the KK gravitons are bound in the gravitational 
potential of a proto-neutron star as it evolves into a neutron star, then the flux of photons from KK graviton decays, 
$G_{KK} \rightarrow 2 \gamma$, could be used to set a limit on extra dimensions. \ They use EGRET results to set limits on LED.  
%We henceforth refer to Hannestad and Raffelt in Ref. \cite{HR2003} as HR.
 \ However, they do not place direct flux limits on the neutron stars not detected by EGRET.  Rather, they quote their flux upper limit as the 1 yr point-source sensitivity of EGRET for a high latitude point source with a $E^{-2}$ spectrum (see section \ref{sec:ExpMeth}), and derive limits on LED more restrictive than from arguments based on KK graviton emissivities from SN 1987A. \ To obtain upper limits, we follow similar theoretical arguments as HR, but we perform a very different analysis, including spectral corrections and upper limit spectral analysis with Fermi-LAT data, on 6 gamma-ray faint NS.% \ We also present combined limits from the 6 sources, more stringent than the limits from any 

%individual source.   
%Cass\'e et al. tried to improve upon the bounds by these authors by combining sources located in the galactic center. \cite{Casse2004}.   

\section{Data Analysis}

\subsection{Experimental Methods\label{sec:ExpMeth}}
The Fermi-LAT is a gamma-ray imaging pair-conversion  telescope, consisting of an anti-coincidence detector, tracker, calorimeter, and electronics modules. %The anti-coincidence detector on the top and wrapping partially around the sides of the Fermi-LAT, is intended for charged-particle background rejection.  The tracker consists of 18 layers, with alternating planes of tungsten foil for pair conversion and silicon-strip detectors for position measurement.  The calorimeter mainly records energy deposition with a hodoscopic array of CsI(Tl) crystals.  The Fermi-LAT has a modular design, and is comprised of a 4$\times$4 array of tower modules, with each module having a tracker module and calorimeter module. 
The details of the Fermi-LAT are discussed by Atwood \emph{et al.}\cite{Atwood_LAT_2009}. The Fermi-LAT features improved performance compared to its predecessor $\gamma$-ray observatory, EGRET.  Some of these specifications, relevant to this study, are compared in Table \ref{tab:compareSpec}.

\begin{table}
\begin{center}
\begin{tabular}{lcc}
\hline
specification & Fermi-LAT & EGRET  \\
\hline
68\% containment PSF ($^\circ$) at 200 MeV& 2.8 & 3.3 \\
Effective Area (cm$^2$) at 200 MeV & 3000 & 1000\\
%FOV (sr) & 2.4 & 0.5 \\
Energy Resolution (\%) at 200 MeV & 13 & 9.3 \\
flux sensitivity (cm$^{-2}$s$^{-1}$) & $6\times10^{-9}$ & $1.3\times10^{-7}$ \\
\hline
\end{tabular}
\end{center}
\caption{Comparison of performance specifications of Fermi-LAT and EGRET, as relevant for the gamma-ray energies considered in this paper\cite{Atwood_LAT_2009,thompson1993}.  Flux sensitivity is evaluated for a high-latitude point source with a $E^{-2}$ spectrum, with 1 year of data, for $E>100$ MeV.  EGRET effective area is quoted for Class A events.}\label{tab:compareSpec}
\end{table}
% specifically, the Fermi-LAT features a narrower PSF at 100 MeV of 3.5$^\circ$ compared to 5.6$^\circ$, a larger field of view of 2.4 sr compared to ~0.5 sr, and an enhanced point source sensitivity for a high-latitude $E^{-2}$ spectrum of 6$\times10^{-9}$ cm$^{-2}$s$^{-1}$ compared to $~10^{-7}$ cm$^{-2}$s$^{-1}$.   
%In addition, deadtime is significantly lower and the effective area is larger.

Setting flux limits on sources requires knowledge of background point sources and diffuse emission.  We make use of the publicly available diffuse models developed by the Fermi-LAT collaboration:  the Galactic diffuse emission model, \emph{gll\_iem\_v02.fit}\cite{Strong2004,Strong2007}; and the isotropic model, \emph{isotropic\_iem\_v02.txt}\cite{isotropicDiff}. 
% The isotropic model is based on spectral line surveys of HI and CO (as a HII tracer) as well as infrared tracers of dust column density.  The $\gamma$-ray emissivities of Galactocentric rings in several energy bands are fit to Fermi-LAT observations, given an model of inverse Compton emission that is calculated using GALPROP \cite{Strong2004} \cite{Strong2006}. 
The Galactic diffuse model is allowed to vary in a region of interest (ROI) around each source by multiplying by a power-law spectral function, as described in \cite{1FGLcatalog}, effectively making the spectrum harder or softer.  The scale for the power-law is 100 MeV, and the index is allowed to vary between -0.1 and 0.1 (a value of 0 implies no correction to the model).   
% is determined as the residual of a fit of the Galactic diffuse emission model to the Fermi-LAT data at Galactic latitudes for $|b|>30^\circ$; it includes contribution from residual cosmic rays at high energies.   
The background point sources are modeled by fixing the spectral parameters from the first year Fermi-LAT (1FGL) catalog \cite{1FGLcatalog}.

	The data sample consists of a selection of 11 months of all-sky data obtained with the Fermi-LAT instrument, using a time interval beginning with the start of survey mode, August 4, 2008, until July 4, 2009. \ This time interval is chosen to be consistent with the 1FGL catalog, so that nearby point sources detected with high significance may be modeled appropriately as power-law sources {\cite{1FGLcatalog}}.  The instrument response function (IRF) chosen is \emph{P6\_V3\_DIFFUSE}\cite{postLaunchPerf}, as is the case for the 1FGL catalog.  This IRF specifies a parametrization of effective area, energy resolution, and point spread function.  We select data from the 1FGL catalog dataset for regions of interest (ROIs) corresponding to each source described further in this paper.     This dataset excludes events for which the rocking angle is larger than 43$^\circ$, because of contamination from the Earth's limb due to interactions of cosmic rays with Earth's upper atmosphere.  For the same reason, for each ROI, events for which the zenith angle is larger than 105$^\circ$ are excluded.  There is also a good time interval (GTI) cut applied, as described in Ref. \cite{1FGLcatalog}.

\subsection{Selection of Neutron Stars}

A query is made on the Australia Telescope National Facility (ATNF) radio pulsar catalog to select NS \cite{ATNFcatalog}.  In order to obtain the best limits, we reject candidate sources that have associations with $\gamma$-ray sources detected in the 1FGL catalog.  To minimize attenuation of the putative KK graviton decay signal, as discussed in section IV, we choose NS that satisfy the following criteria: distance $d<0.40$ kpc; surface magnetic field $B_{\tmop{surf}}<5 \times 10^{13}$ G; and characteristic age $t_{\tmop{age}}< 2\times 10^8$ yr.  We take the NS ages as the spin-down ages of the pulsars, for consistency over all sources; the corrected ages may differ, as discussed in Ref. \cite{bijanThesis}.   However, consideration of the corrected ages hardly affects the limits presented here.  In addition, the $\gamma$-ray sky as viewed by Fermi-LAT is filled with sources near the Galactic plane, and diffuse components are also dominant and have large systematic uncertainties at low latitudes;  we require for Galactic latitude ($b$), that $|b| > 15^{\circ}$ for candidate neutron stars. Applying all the above selection criteria to the ATNF catalog, 6 sources remain for analysis, with parameters as shown in Table \ref{tab:srcTable}.% \cite{ATNFcatalog}.
%\footnote{For the source yielding the best limits, RX~J1856$-$3754, the corrected age is 3.79 Myr, whereas the the spin-down age is 3.76 Myr}.  %The criterion on the distance is the single most important factor, because distance ($d$) degrades the flux limits as $1/d^2$.   
\begin{table}
\begin{centering}
\small
%\newcolumntype {d }[1]{ D { ,}{.}{#1}}
\begin{tabular}{|l|d{2}|d{2}|d{2}|d{2}|d{2}|l|l|l|}
\hline
source name & \mbox{RA} & \mbox{Dec.}  & \mbox{$\ell$}  & \mbox{$b$}  & \mbox{$P$} & $d$  & Age & $B_{\mbox{surf}}$ \\
 & \mbox{($^\circ$)} & \mbox{($^\circ$)} & \mbox{($^\circ$)} & \mbox{($^\circ$)} & \mbox{(s)} & (kpc) & (Myr) & (G) \\
\hline
RX~J1856$-$3754 & 284.15 & -37.90 & 358.61 & -17.21 & 7.05 & 0.16 & 3.76 & 1.47$\times 10^{13}$ \\
\hline
J0108$-$1431 & 17.04 & -14.35 & 140.93 & -76.82 & 0.808 & 0.24 & 166 & 2.52$\times 10^{11}$ \\
\hline
J0953$+$0755 & 148.29 & 7.93 & 228.91 & 43.7 & 0.25 & 0.26 & 17.5  & 2.44$\times 10^{11}$ \\
\hline
J0630$-$2834 & 97.71 & -28.58 & 236.95 & -16.76 & 1.24 & 0.33 & 2.77& 3.01$\times 10^{12}$ \\
\hline
J1136$+$1551 & 174.01 & 15.85 & 241.90 & 69.20 & 1.19 & 0.36 & 5.04 & 2.13$\times 10^{12}$ \\
\hline
J0826$+$2637 & 126.71 & 26.62 & 196.96 & 31.74 & 0.53 & 0.36 & 4.92 & 9.64$\times 10^{11}$ \\
\hline
\end{tabular}
\end{centering}
\caption[justification=justified]{Astrophysical properties of neutron star sources analyzed in this work, with sources in increasing order of distance. Coordinates, periods. distances, ages, and surface magnetic field strengths are obtained from the ATNF Catalog \cite{ATNFcatalog}.}
\label{tab:srcTable}
%\end{centering}
\end{table}

\subsection{Gamma-ray Spectral Limits \label{sec:roi}}
\begin{figure}
\begin{center}
%\begin{array}{ll}
\includegraphics[clip=true,trim=0.06in 0in 0.02in 0in,width=3.36in, keepaspectratio=true]%
{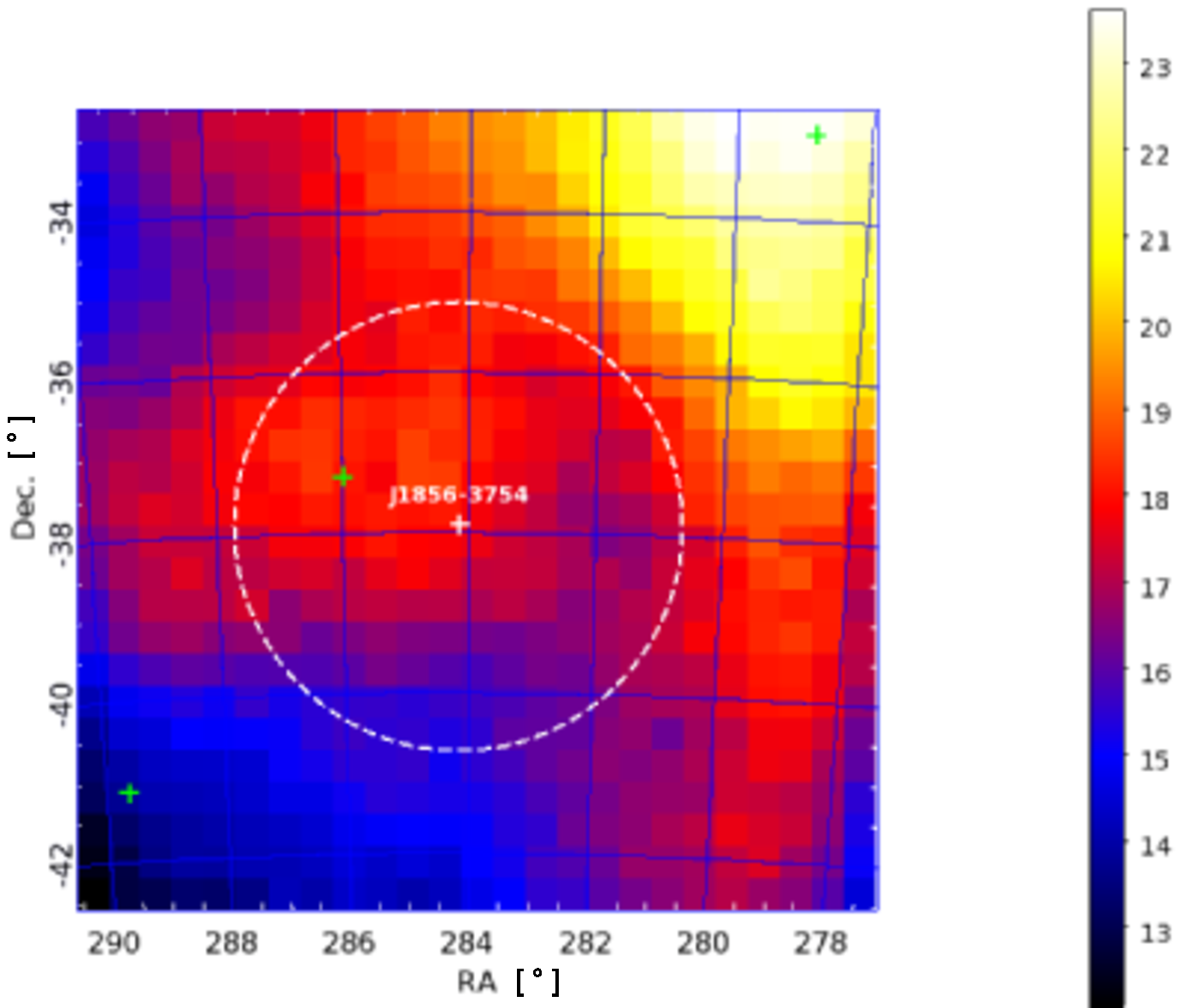}%{cmap_100-400MeV_front_back_fgauss_add_10degSq_crop_ds9_win_SLAC}
\includegraphics[clip=true,trim=0.06in 0in 0.02in 0in, width=3.36in,keepaspectratio=true]%
{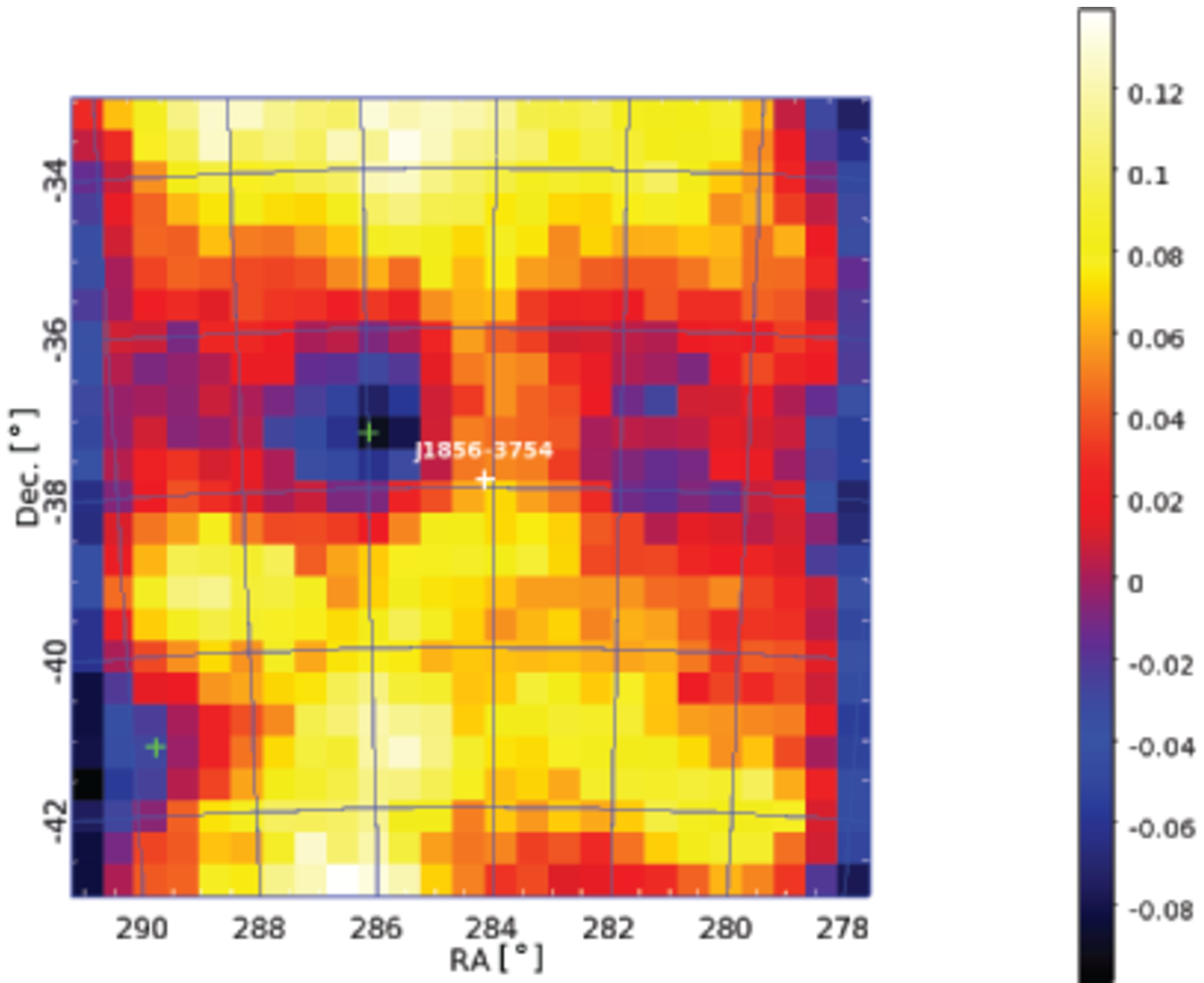}
%\subfloat[Model Map.]{\includegraphics[width=3.25in]{mdlMap_J1856_pnm}}

\caption{\label{fig:ROI_J18561}Left: Counts map from data, for source RX~J1856$-$3754, convolved with a Gaussian approximation to the Fermi-LAT PSF, in order to reduce statistical fluctuations without dramatically reducing the angular resolution.  The colorbar shows counts per pixel.  The white dashed circle shows the 200 MeV PSF.  Right: Residual map, (counts-model)/model, for the same source, based on the 1FGL model with the fitted diffuse model.  The pixel size is $0.4^\circ$ for both.  Green crosses show 1FGL point sources, and the putative $\gamma$-ray source is at the center.}  %The linear color-scale shown is the same for both maps.}
\end{center}
\end{figure}

\begin{figure}
\centering
\includegraphics[trim=0in 0.3in 0in 0.2in,width=4in]{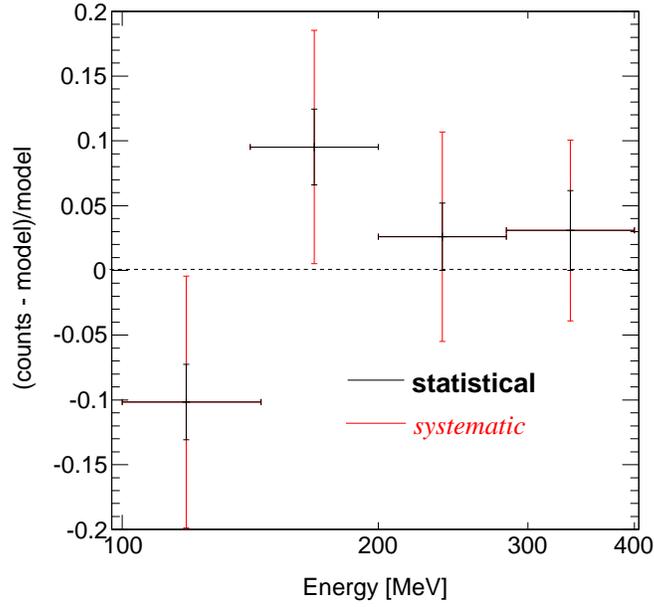}
\caption{\label{fig:ROI_J1856_resid}  Residual plot of counts over the 10$^\circ$ square region around source RX~J1856$-$3754.  Black horizontal bands indicate the energy range for each point, and black vertical bands represent statistical uncertainties, while red vertical bands represent systematic uncertainties of about 10\% at 100 MeV, and decreasing to 5\% at 562 MeV\cite{fermiDiffuse2009}.}
\end{figure}

Fermi-LAT gamma-ray events are selected in a 12$^\circ$ radius ROI centered on each NS source listed in Table \ref{tab:srcTable}.  Given the limitations of the dataset we use and the 
expected spectral energy distribution from gamma rays from trapped KK graviton decay, only gamma-ray events with energies in the range 100 MeV to 400 MeV are considered.  Although desirable, going below 100 MeV is not feasible for this analysis using \emph{P6\_V3\_DIFFUSE}.  Before obtaining upper limits for each source, a model for the corresponding ROI is developed, inclusive of 1FGL sources and the 2 components of diffuse emission (no putative neutron star source is included in this step).  1FGL catalog sources within a 14$^\circ$ radius are parametrized as point sources with a power-law spectral energy distribution, with fluxes and spectral indices \emph{fixed} at catalog values; those sources farther than 14$^\circ$ away are not considered.     
These parameters are fixed due to the small gamma-ray energy range (100 MeV-400 MeV) that we use, which leaves little spectral range to perform accurate fitting. \ \ %Two components of diffuse emission are accounted for:  Galactic diffuse and isotropic extraGalactic diffuse.  
An initial unbinned likelihood fit is done in order to determine \emph{only} the diffuse parameters.  For the isotropic diffuse component, the parameter to be determined is normalization, while for the Galactic diffuse component, we consider the normalization and the spectral index.
	The analysis, including the diffuse fitting and upper limit determination, is performed with the Fermi \emph{ScienceTools} program \texttt{pyLikelihood}, featuring maximum likelihood-based fitting. Version \emph{09-17-00} of the Fermi-LAT \emph{ScienceTools} is used\cite{Cicerone}.  For the neutron star RX~J1856$-$3754, a counts map of 100-400 MeV photons in a $10^\circ\times10^\circ$ region, convolved with a Gaussian approximation to the Fermi-LAT PSF, is shown in the left panel of Figure \ref{fig:ROI_J18561}.  The residual counts map (determined from comparison of the counts map to the model-based map), for source RX~J1856$-$3754, is displayed in the right panel of Figure \ref{fig:ROI_J18561}.  Figure \ref{fig:ROI_J1856_resid} shows a residual counts plot versus energy, obtained by integrating over the counts map spatial dependence, and subtracting the data counts from model counts and dividing by the model counts.  %computed from the total counts, including background point sources, the fitted diffuse model, and model predicted counts, is shown in Figure

%\begin{comment}
%\end{comment} 

%\section{Theoretical Modeling of the KK Graviton Decay $\gamma-$Ray Spectrum}  

%\section{Developing the Spectral Energy Distribution by Theoretical Simulation}
%Obtaining Limits on LED from KK Graviton Decay}

A spectral model, which determines the differential flux $d\Phi/dE$, for each source and number of extra dimensions $n$, is developed in the next section. A significant difference in the data analysis technique from Hannestad and Raffelt lies in considering the differential flux rather than the integral flux, in determining limits on $R$; this is a more accurate and optimal method in setting limits in a Fermi-LAT analysis, when comparing the data to a pre-defined theoretical distribution.  Complete details of the theoretical development of the differential flux, as well as analysis methods, can be found in \cite{bijanThesis}.    %$distance $d$ in kiloparsecs, extra dimensions size $R$ in meters, $n$, the unique distribution $dN_{n}/dE$, and an $n$-dependent constant arising from theory,  applicable for all sources, is given as:

%Values for $k_n$ are shown in Table \ref{tab:k_n_table}. 
%The distribution $dN_n/dE$ for each candidate source is developed as a histogram from Monte Carlo simulation as a function of $B_{\tmop{surf}}$, and $t_{\tmop{age}}$ for different $n$ as described in the following subsections of this article; representative distributions are shown in Figure \ref{fig:dI_dE}.  %Then, to facilitate the analysis, each is parametrized in terms of a power-law super-exponential cutoff function, 

%\begin{equation} \frac{dN_n}{dE} = N_0 \left(\frac{E}{E_0}\right)^{\gamma_1}\exp\left(-\left(\frac{E}{E_c}\right)^{\gamma_2}\right) \end{equation} where $N_0$ is the normalization prefactor, $E_0= 100$MeV is the fixed scale, $E_c$ is the cutoff, $\gamma_{1}$ is the spectral index of the power-law, and $\gamma_2$ is the index of the exponential cutoff. 

%
\section{Calculating the Spectral Model for KK Graviton Decay $\gamma-$Rays from NS}

In the following subsections, we explain how we calculate the gamma-ray spectrum.  Important departures from the analysis of HR in forming the theoretical differential flux, $d\Phi/dE$, to be compared to Fermi-LAT observations include: attenuation of the signal due to the age of the neutron star, orbital position and velocity of the $G_{KK}$, decay of $G_{KK} \to 2\gamma$, and attenuation of the signal due to magnetic field (which is position and velocity dependent).  These features are included via a Monte Carlo simulation of about $10^7\ G_{KK}$ in orbit for each NS source and for $n=2,3,\ldots,7$ extra dimensions.  %In the following subsections, the development of the gamma-ray spectrum $dN_n/dE$ is explained.  Important departures from the theoretical formalism of Hannestad and Raffelt include the relativistic energy distribution of the gravitons, correction to the gamma-ray energy distributions due to the ages and magnetic fields of neutron stars, and consideration of the radial spatial distributions around the neutron stars due to the orbits of the gravitons.  The relativistic energy distribution of the gravitons is determined from the mass distribution and the the velocity distribution of the gravitons.  The mass distribution, in turn, is determined from sampling theoretically-motivated functions for the gravitons' kinetic energy and Lorentz parameter.  The gravitons' velocity distribution is given by the time derivative of the gravitons' orbital trajectory.  Approximately $10^{7}$ gravitons are simulated in determining the spectral energy distribution for each source and each number of additional spatial dimensions.  
%The simulation is performed in C++ code, gcc version\cite{gcc}, and is compiled against libraries in the high-energy physics analysis framework ROOT, version 5.26\cite{ROOT}, as well as GNU Scientific Libraries (GSL), for the use of special functions, such as Bessel functions\cite{gsl}.  %Attenuation effects are treated as follows.  
%There is a survival probability  $P_1$ of a simulated event due to attenuation from the $h-$decay, and a survival probability $P_2$ of an event due to attenuation from pair production in the NS magnetosphere.  Simulated events are rejected if a random uniformly-generated numbers $u_1, u_2$ on the interval (0,1) satisfy: 
%\begin{equation} u_1 > P_1 , u_2 > P_2 \end{equation}.  
%The random generator used is the Mersenne-Twister algorithm (as/Table implemented in ROOT by the class \texttt{TRandom})\cite{refStat},\cite{ROOT}.

\subsection{Theoretical Model}\label{sec:TheoreticalModel}

Following HR, we start with the differential distribution of $G_{KK}$ created during the proto-neutron star core collapse with total energy $\omega$ and  mass $m$:
\begin{equation}
\frac{\dd^2 N_{\rm KK,n}}{\dd\omega\dd\mu} = \frac{\dd^2 Q_n}{\omega\dd\omega\dd\mu}\Delta t_{NS}V_{NS}\ .\label{eqn:dNKK}
\end{equation}

\noindent 

%Throughout this paper, 
We will make use of notations defined in HR, that we rewrite here for completeness: $Q_n$ is the total energy loss rate per unit volume into \gravitons\, which depends on $n$; $\mu=m/\omega$ is the inverse of the initial Lorentz factor of the $G_{KK}$; $\Delta t_{NS}\simeq7.5$ s is the time-scale for emission of $G_{KK}$ during the core collapse; and $V_{NS}=\frac{4}{3}\pi R_{NS}^3$ is the volume of the proto-neutron star (and current neutron star) of radius $R_{NS}\simeq 13$ km.  According to HR, we have: 
\begin{equation}\label{eqn:diffQ1}
 \frac{\dd^2 Q_n}{\dd\omega\dd\mu} = Q_0(RT)^n \Omega_n G_{n-1}(\mu)F_{n}\left(\frac{\omega}{T}\right)\ ,
\end{equation}
\noindent where $R$ is the extra dimension size, as in eq. (1), $T\gtrsim30$ MeV is the supernova core temperature (see Section \ref{sec:int_and_concl}), and $\Omega_n=2\pi^{n/2}/\Gamma(n/2)$ is the surface of the $n$-dimensional unit hypersphere, with $\Gamma(...)$ as the Gamma function.  We also have,

%The constant $S_0$ is defined as:
%\begin{equation} S_0 = \frac{1024\pi^{1/2}}{5}G_N\sigma n_B^2 T^{5/2} M^{-1/2} \end{equation}
\begin{align} 
{Q_0} &= \frac{{512}}{{5{\pi ^{3/2}}}}\frac{G_N \sigma n_B^2 T^{7/2}}{M^{1/2}} \\
 &= 1.100 \times {10^{22}}{\rm{ MeV  }} \ {{\rm{cm}}^{ - 3}}{\rm{ }}{{\rm{s}}^{ - 1}} \left(T/30 \ {\rm MeV}\right)^{7/2} \left(\rho/3\times10^{14} \ {\rm g \ cm}^{-3}\right)^2  (f_{KK}/0.0075)\ ,\end{align}

\noindent where Newton's constant is $G_N=6.708 \times 10^{-33} \hbar c\, ({\rm MeV}/c^2)^{-2}$, $\sigma$ is the nucleon-nucleon scattering cross section of 25 mb, $n_B$ is the number density of baryons of 0.16 fm$^{-3}$, and $M$ is the isospin-averaged nucleon mass of 938 MeV/$c^2$.  $f_{KK}\simeq0.01$ is the estimated fraction of core-collapse energy radiated away as $G_{KK}$\cite{egretDiffuse,HR_SN_diffuse}. 
In eq. (\ref{eqn:diffQ1}), the following functions are defined, where $q$ is an integer:

\begin{align}
 G_{q}(\mu) &= \mu^{q}\sqrt{1-\mu^2}\left(\frac{19}{18}+\frac{11}{9}\mu^2+\frac{2}{9}\mu^4\right) \\
 F_{q}(\omega/T) &= \frac{(\omega/T)^q}{1+\exp(\omega/T)}.
\end{align}

In the previous equation, in writing $F(\omega/T)$, we are assuming that the structure function of the nuclear medium, in the notation of HR, $s(\omega/T)$, of the nuclear medium is unity, which is accurate to first order\cite{HR2003}.  An expansion to the next order, $(\omega/T)^2$, would likely shift the expected energy distribution of the differential flux to higher energies. Therefore, our assumption of $s(\omega/T)\simeq 1$ is in the direction of making the associated limits more conservative.
%Next, we define the differential decay rate into a pair of gammas at the present epoch :
%\begin{equation}\label{eqn:diffQ2}
% \frac{\dd N_{\gamma}}{\dd\omega\dd\mu\dd t} = n_{\rm KK}(\omega,\mu) \times \frac{2}{3}\frac{\mu}{\tau(m)}\exp\left(-\frac{\mu\, t_{age}}{\tau(m)}\right)\ ,
%\end{equation}
%where $2/3$ is the $KK\rightarrow \gamma\gamma$ branching ratio multiplied by the final state photon multiplicity, $\tau(m)$ 
%is the rest frame lifetime of the \graviton, and $t_{age}$ the age of the neutron star. While HR set the exponential term to 1, 
%a reasonable assumption /given that $t_{age}$ is typically much smaller than $\tau(~100{\rm MeV})$, the Monte-Carlo procedure that we
%describe in the next section allows us to do away with this assumption, whence the presence of this term in eq.~(\ref{eqn:diffQ2}).
Finally, the integral for the case of trapped KK gravitons, which make up the initial cloud bound to the NS, is given by: % where $g_{KK}$ creation occurs, can be expressed as,

\begin{equation}
 N_{KK,n}(t=0) = 3 \int_0^\infty d\omega \ \int_0^1 r^2 \ dr  \int_{1 + U (r)}^1 d\mu \ \frac{\dd^2 N_{KK,n}}{\dd\omega\dd\mu}. \label{eqn:N_KK_n_0}
\end{equation}

In eq. (\ref{eqn:N_KK_n_0}), HR assume that the graviton creation is isotropic at the dimensionless radial distance from the neutron star center, $r$, scaled to the neutron star radius, $R_{NS}$.  The integration over $r$ is performed from the proto-neutron star's center to its surface, where $r=1$, and the condition $\mu>1 + U (r)$ selects the $G_{KK}$ that are gravitationally trapped. As in HR, we model the neutron star's potential as Newtonian:
\begin{equation}
U(r) = -\frac{G_{\tmop{N}}M_{\tmop{NS}}}{R_{\tmop{NS}}c^2}\times\left\{  
\begin{array}{ll}
   \left(\frac{3}{2}-\frac{1}{2}r^2\right)\ , & r < 1\\
    \frac{1}{r}\ , & r \geq 1
  \end{array}\right.
\end{equation}
\noindent with $U_{NS} = - G_N M_{NS}/(R_{NS}c^2)=-0.159(M_{NS}/1.4M_\odot)(13{\rm km}/R_{NS})$.  
%\noindent with values tabulated in Table \ref{tab:k_n_table}. 

\noindent The $G_{KK}$ lifetime is \cite{Han1998sg}\footnote{This takes into account competing decays to $e^+e^-$ and $\nu\bar{\nu}$.}

\begin{equation} \tau(m) = 1\times 10^9 \mbox{yr} \left(\frac{100 \mbox{MeV}}{m}\right)^3 \equiv \kappa^{-1}m^{-3}, \label{eqn:tau} \end{equation}

\noindent where  $\kappa = 3.17\times10^{-23} \ \mbox{MeV}^{-3} \mbox{s}^{-1}$.  Assuming an exponential decay of the KK gravitons, the number of KK gravitons remaining at time $t$ after the core collapse is given by:

\begin{equation} N_{KK,n}(t) = N_{KK,n}(t=0)\exp\left(-\frac{\mu t}{\tau(m)}\right) \end{equation}

\noindent Then, the time derivative (absolute value), of eq. (\ref{eqn:dNKK}), is given by:

\begin{equation}
\begin{split} \left|\frac{\dd^2\dot{N}_{KK,n}}{\dd \mu \dd \omega}\right| &= \kappa \frac{m^4}{\omega} N_{KK,n}(t) \\
				%&= \kappa \frac{m^4}{\omega^2}	Q_n \Delta t_{NS} V_{NS} \\
				%&= \kappa \mu^4\omega^2 Q_n \Delta t_{NS} V_{NS} \\ 
				%&= \kappa\Omega_{n} Q_0 \mu^4\omega^2  G_{n-1}(\mu) F_{n}(\omega/T)\Delta t_{NS} V_{NS}\\
				&= Q_0 (RT)^n \Omega_n \Delta t_{NS} V_{NS} \kappa T^2 G_{n+3}(\mu) F_{n+2}(\omega/T) \exp\left(-\frac{\mu t}{\tau(m)}\right). 
                                %\\
			%	&= N_{0,n} (RT)^n \kappa T^2  G_{n+3}(\mu) F_{n+2}(\omega/T) 
\end{split}
\label{eqn:dNdot}
\end{equation}
 %The photon number flux, $\dot{N}_\gamma$, due to KK graviton decay, is related to the number of gravitons that have decayed into gamma rays:%, where there are two photons per gravitons decay and a BR of 1/3 to photons:
%The flux of gamma rays depends upon the decay rate of gravitons, as expressed in eqn. (\ref{eqn:dNdot}).

\subsection{Determining the Differential Flux by Monte Carlo Simulation\label{sec:SED_MC}}
We determine the differential flux according to a Monte Carlo simulation that uses eq. (\ref{eqn:dNdot}).  %, motivated by the preceding equation.  %, in order to determine the spectrum $\dd _n/\dd E_\gamma$, which is proportional to $\dd \Phi/\dd E_\gamma$. 
We calculate the mass distributions and the Lorentz parameters of the decaying gravitons.  We then determine the momentum and energy distributions of the $G_{KK}$, considering the geometry of the decays.  At the same time, the age of the NS determines the remaining number of gravitons.  We also consider whether a given gamma ray can escape the NS magnetosphere.  Finally, this determines the differential flux of gamma rays from the NS.

We carry out the Monte Carlo simulation of the differential flux in the following steps:

\begin{enumerate}[1]
\item) Sample $\omega$ from $F_{n+2}(\omega/T)$, as in eq. (\ref{eqn:dNdot}), for $0 < \omega/T < 20$ ($T=30$ MeV).  
For $\omega/T>20$, there is negligible contribution from the integral of $F_{n+2}(\omega/T)$.
\item) Sample $\mu$ from $G_{n+3}(\mu)$, as in eq. (\ref{eqn:dNdot}), between $\mu_{\min}$ and $\mu_{\max}$.
(Note that the sampling steps 1 \& 2 are independent of each other, see HR.)  
To simplify the orbit calculation with only a small error, we assume that all of the created $G_{KK}$ start their orbit at the center of the NS ($r=0$)\footnote{We have calculated that this approximation is in the direction of making our limits more conservative.}. Thus we have $ \mu_{\min}=0.807$,  corresponding to the $G_{KK}$ escape velocity, and $\mu_{\max}=0.926$, corresponding to the minimum velocity to reach the neutron star surface, from $r=0$.  Having determined a value of $\mu$, we determine the initial $G_{KK}$ Lorentz factor $\gamma=1/\mu$ and initial velocity $\beta=\sqrt{1-\mu^2}$. Given the geometry of the SN explosion, we assume, as do HR, that the $G_{KK}$ orbits are radial.  Using $\omega$ and $\mu$, from steps 1 and 2, we determine a value for the mass, $m=\mu\omega$.  Representative distributions of $m$, for different values of $n$, are shown in Figure \ref{fig:hMass_n257}.% 

\begin{figure}
\begin{centering}
\includegraphics[width=0.9\textwidth]{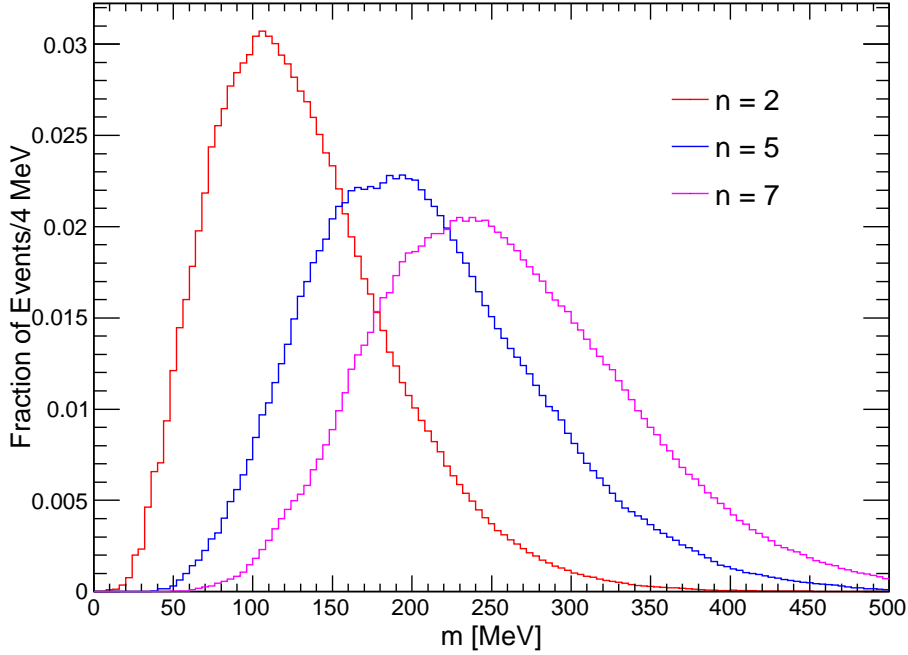}
\caption{\label{fig:hMass_n257} Unit-normalized distributions of KK graviton masses for $n = 2,5,7$, as determined according to Monte Carlo simulation.}
\end{centering}
\end{figure}

\indent	Since we know the mass at this point, and given the age of the neutron star, $t_{\tmop{age}}$, we calculate the exponential decay fraction, $F_{\tmop{decay}} = \exp\left(-\frac{\mu t_{\tmop{age}}}{\tau(m)}\right)$.  We sample a real number $u$ uniformly in the interval [0,1]: if $u>F_{\tmop{decay}}$, then the event is rejected.

\item) Sample the decay vertex $r_0$ for a given $\mu$.  
%We sample $r_0$ for $0 < r_0 < r_{\max}$ ($r_{\max}$ is defined in eq. (\ref{eqn:rmax})), but reject $r_0$ if it is less than 1, due to the opacity of the neutron star matter to gamma-rays.  
The probability density function $P(r_0;\mu)$, which is shown in Figure \ref{fig:RadialDistributions} for two values of $\mu$, is obtained as described in Appendix \ref{sec:appendixA}.
In Figure \ref{fig:radialProfile}, unit-normalized radial profile of decay vertices as a function of radial coordinate, for n = 2, is plotted. 
%On the y-axis is plotted $\left<P(r; \mu)\right>\mu$, averaged over all $\mu$ between $\mu_\min$ and $\mu_\max$, for 1 < r < 5.5.

%a probability density function for $r$, given a value of $\mu$; examples are shown in Figure \ref{fig:RadialDistributions}. 

%The full orbit cycle would include a period from $0<r<r_{\max}$, passing through the center of the neutron star to the other side; thus our distribution is taken over a quarter of an orbit cycle.  However, due to the symmetry of the orbit about $r=0$ and $r=r_{\max}$, this treatment of sampling $r_0$ is valid.

\begin{figure}
  \centering
  \subfloat[$\mu = 0.83 $]{\label{fig:hRadial_10}\includegraphics[width=0.50\textwidth]{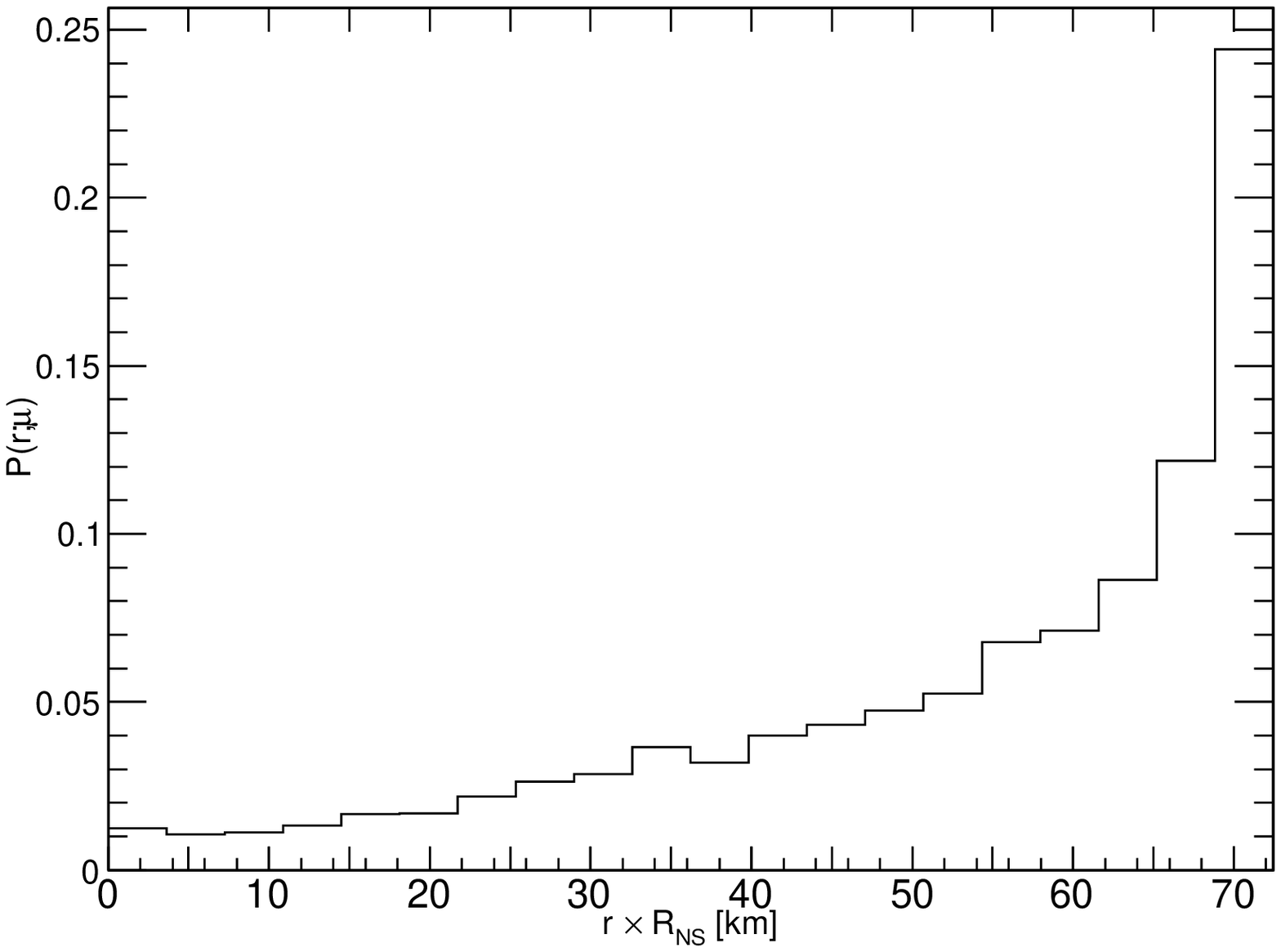}}                
  \subfloat[$\mu  = 0.90$]{\label{fig:hRadial_40}\includegraphics[width=0.50\textwidth]{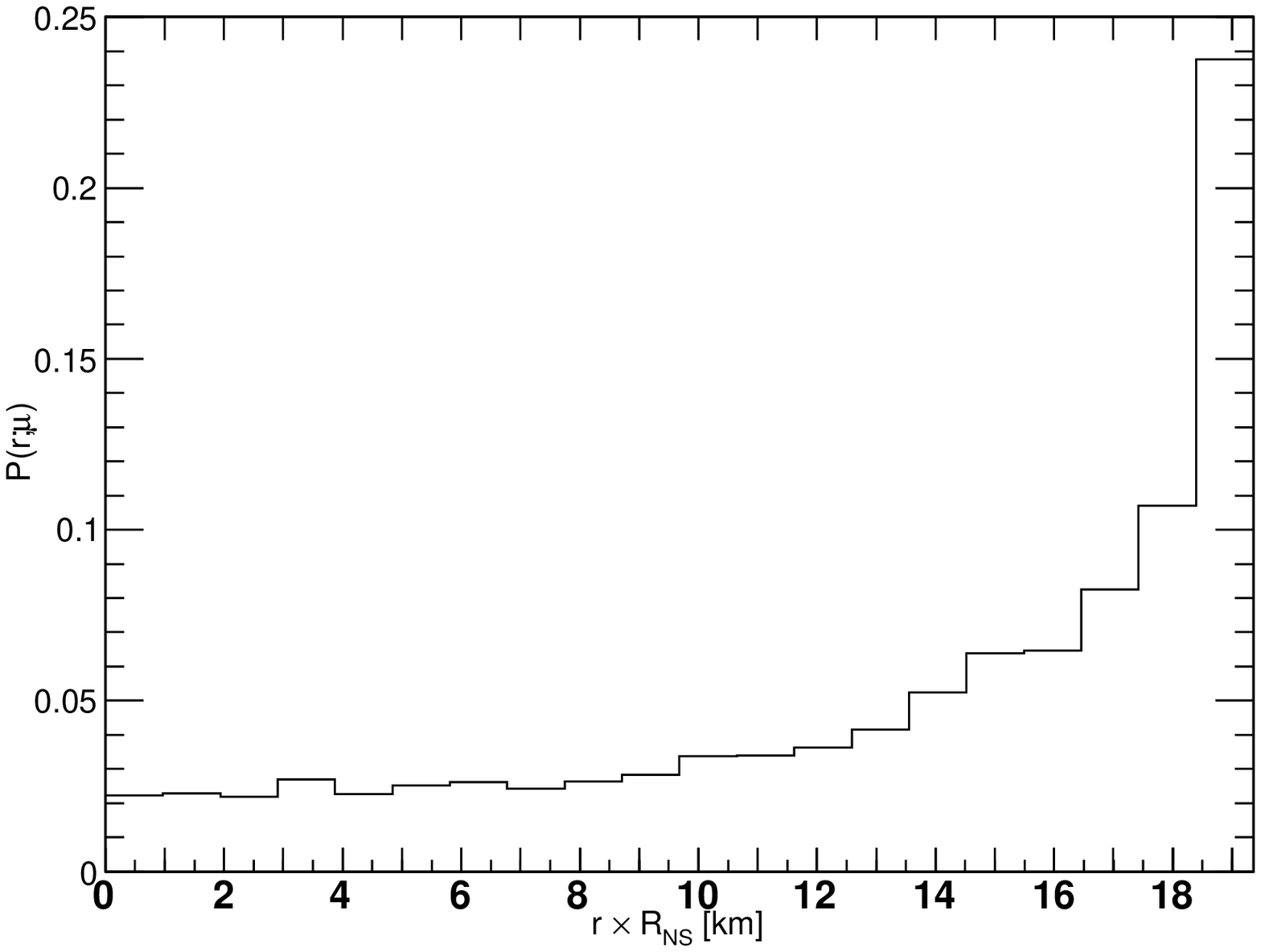}}
  \caption{Radial probability density functions, $P(r;\mu)$, for 2 different values of $\mu$.  There are 20 linearly-spaced bins over the interval $[0,r_{\max}\times R_{NS}]$, and the $y-$axis is the fraction of events per bin.}
  \label{fig:RadialDistributions}
\end{figure}

\begin{figure}
\begin{centering}
\includegraphics[width=5in]{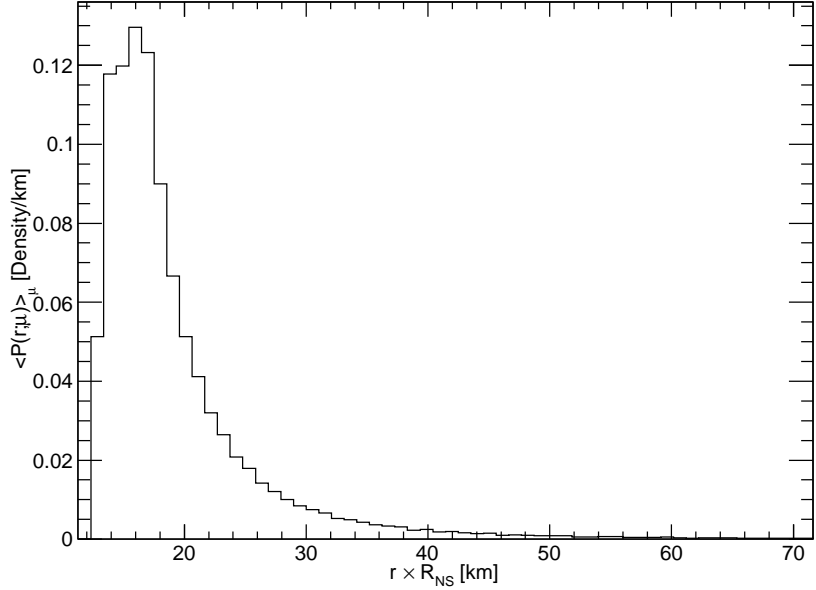}
\caption{Unit-normalized radial distribution of decay vertices as a function of radial coordinate, $r_{\tmop{km}}$, for $n=2$.  On the y-axis is plotted $\left<P(r;\mu)\right>_\mu$, averaged over all $\mu$ between $\mu_{\min}$ and $\mu_{\max}$, for $1<r<5.5$.}\label{fig:radialProfile}
\end{centering}
\end{figure}

\begin{comment} 
\begin{figure}
  \centering
  \subfloat[$\beta = 0.4$]{\label{fig:beta0.4}\includegraphics[width=6in]{trajectory2cOut_betaArg_4}}     
\subfloat[$\beta  = 0.5$]{\label{fig:beta0.5}\includegraphics[width=6in]{trajectory2cOut_betaArg_5}}
  \caption{Trajectories for 2 different values of $\beta$, $R$ (m) vs $t$ (s).}
  \label{fig:TrajectoriesBeta1}
\end{figure}
\end{comment}

\item) Sample the orbit direction isotropically ($-1 < \cos\theta< 1$, $-\pi<\phi<\pi$).  This selects an orbit direction:

\begin{equation} \hat{r}_0=\sin\theta\cos\phi \ \hat{x}+\sin\theta\sin\phi \ \hat{y}+\cos\theta \ \hat{z} \end{equation}

where $\hat{x},\hat{y},\hat{z}$ are the unit coordinate directions in the NS frame.  The $\hat{z}$ direction is chosen to align with the magnetic dipole axis of the NS. At the sampled decay vertex, $r_0$, we then obtain a velocity, 

\begin{equation} \beta^\prime=R_{NS}\dot{r}/c, \end{equation}

where $\dot{r}=\left.dr/dt\right|_{r=r_0}$ is obtained numerically, and the Lorentz factor,

\begin{equation} \gamma^\prime = 1/\sqrt{1-\beta^{\prime \ 2}}.\end{equation}

Using the determined values of $m$ and $\beta^\prime$, thus yields % $\vec{\beta}^\prime$, as well as the $G_{KK}$ momentum in the NS frame, 
$p_{KK}^\nu$, the $G_{KK}$ 4-momentum in the NS frame at the decay vertex.

\item) Determine the energy and momentum distribution of one of the two decay photons at decay point on the orbit with direction $\hat{r}_0$.  
% $\vec{r}_0=\hat{r}_0r$. 
We treat the other decay photon by multiplying the final flux by two.  Full details of this procedure, as implemented in the Monte Carlo simulation, are given in Appendix \ref{sec:appendixB}. %, given \ 

\item) Determine whether the photon pair-produces in the neutron star magnetosphere: this probability is given by $P_{\tmop{pp}}(E_\gamma,\vec{r},\vec{p}_\gamma)$.  The probability for photon survival from pair production in the Monte Carlo simulation is then taken as $P_{\tmop{pp}}(E_\gamma,\vec{r},\vec{p}_\gamma)=\exp(-\tau_{pp})$.  In Appendix \ref{sec:appendixC}, we describe the computation of $\tau_{pp}$.  We sample a real number $v$ uniformly on the interval [0,1]: if $v>P_{\tmop{pp}}$, then the event is rejected. 
% The direction and energy of each photon, determined in the previous steps, are important in determining the pair production attenuation probability, as discussed in the next subsection.  If a photon has a decay vertex of $r\geq7$, do not evaluate $P_{\mbox{pp}}$ and accept the event, because the magnetic field is 0.3\% of the surface field.  

\end{enumerate}
%\end{indent}
%\bullet //
\subsection{The Final Flux Result from the Monte Carlo Simulation}

The distribution defined by the Monte Carlo simulation, $dN_n/dE_\gamma$, is related to the differential flux by:

\begin{equation}\frac{{d\Phi }}{{dE_\gamma}} = {k_n}{R^n}(d/{\rm kpc})^{ - 2}\frac{dN_n}{dE_\gamma},\end{equation}

\noindent where the $n-$dependent constant $k_n$ is given as:

\begin{equation}{k_n} = \frac{1}{{4\pi {{(3.086 \times {{10}^{21}})}}}^2}T^2 \kappa \left(\frac{T}{\hbar c}\right)^n \frac{2}{3} N_{0,n} \ {\rm{ c}}{{\rm{m}}^{ - 2}}{{\rm{s}}^{ - 1}}{{\rm{m}}^{ - n}}. \label{eqn:k_n}\end{equation}

%The factor of $2/3$ arises from 2 photons per graviton decay, and a branching ratio into photons of 1/3 (see below).  
and
\begin{equation} N_{0,n} = N_{KK,n}(t=0)/(RT)^n \label{eqn:N_0n}, \end{equation}

\noindent where the factor $\kappa$ is related to the decay rate as in eq. (\ref{eqn:tau}), and the factor $T/(\hbar c)$ is a conversion constant, which is numerically $1.52033\times10^{14} \ \mbox{m}^{-1}$ at $T=30$ MeV.  Values of $N_{0,n}$ and $k_n$ are tabulated in Table \ref{tab:k_n_table}. 

In the computation of $dN_n/dE_\gamma$, steps (3) and (7) of Section \ref{sec:SED_MC} reject events based on the decay from the lifetime and the pair production optical depth, respectively.  In the case of a zero-age, zero-magnetic field neutron star source, the spectrum $dN_n/dE_\gamma$ is normalized to 1.  Formally, the distribution $dN_n/dE_\gamma$ is defined by:

\begin{equation} \frac{dN_n}{dE_\gamma} = \frac{1}{N_{ev}}\frac{dN_n^\prime}{dE_\gamma} \label{eqn:dNdE} \end{equation}

\noindent where $N_{ev}$ is the number of events in the Monte Carlo simulation.  $N_{rem}$, the number of events remaining after the effects of decay and pair production are taken into account, is given by the integral:
\begin{equation} N_{rem} = \int_0^{600 \tmop{MeV}} \frac{dN_n^\prime}{dE_\gamma} dE_\gamma. \end{equation}
The upper limit of 600 MeV is determined by the condition $\omega/T\leq20$\footnote{The range of gamma-ray energies used to generate $dN_n^\prime/dE$ in the Monte Carlo is $0<E_\gamma<600$ MeV.}.
% where the fit is used to extrapolate the spectrum from $E_\gamma=100$ MeV to $0$ MeV.  
\noindent The parameter $\eta$, defined as,  

\begin{equation} \eta \equiv \frachoriz{\int_{100 \ {\rm MeV}}^{400 \ {\rm MeV}}\frac{dN_n}{dE_\gamma} \dd E_\gamma}{\int_{100 \ {\rm MeV}}^{400 \ {\rm MeV}} \left. \frac{dN_n}{dE_\gamma}\right|_{\rm non-atten}\dd E_\gamma}. \label{eqn:etaDef} \end{equation}

%\begin{equation} \eta \equiv \frac{\int_{100 {\rm MeV}}^{400 {\rm MeV}} \frac{dN_n}{dE_\gamma} \dd E_\gamma}{\int_{100 {\rm MeV}}^{400 {\rm MeV}}  \frac{dN_n}{dE_\gamma}|_{\rm non-atten} \dd E_\gamma }, \label{eqn:etaDef}\end{equation} 
%\begin{equation}\eta = N_{rem}/N_{ev}, \label{eqn:etaDef}\end{equation}
%\begin{equation} 
\noindent parameterizes the efficiency with which photons contribute to the spectrum,  after signal attenuation effects of lifetime and pair production have been taken into account.   Values for each source and $n$ are shown in Table \ref{tab:etaPar}.

\begin{table}
\begin{centering}
\begin{tabular}{|l|l|l|}
\hline
$n$ & $N_{0,n}$ & $k_n ({\rm{ c}}{{\rm{m}}^{ - 2}}{{\rm{s}}^{ - 1}}{{\rm{m}}^{ - n}})$  \\
\hline
2 & 6.47$\times 10^{40}$ & 7.126$\times 10^{6}$ \\
\hline
3 & 3.46$\times 10^{41}$ & 5.799$\times 10^{21}$ \\
\hline
4 & 1.94$\times 10^{42}$ & 4.963$\times 10^{36}$ \\
\hline
5 & 7.40 $\times 10^{43}$ & 4.511$\times 10^{51}$ \\
\hline
6 & 7.05$\times 10^{43}$ & 4.355$\times 10^{66}$ \\
\hline
7 & 4.97$\times 10^{44}$ & 4.452$\times 10^{81}$ \\
\hline
\end{tabular}
\caption{$n-$dependent constants, as defined in equations (\ref{eqn:N_0n}) and (\ref{eqn:k_n}).}
\label{tab:k_n_table} 
\end{centering}
\end{table}

\begin{table}
\begin{centering}
\small
%\footnotesize
\begin{tabular}{|l|l|l|l|l|l|l|}
\hline
n & RX~J1856$-$3754 & J0108$-$1431 & J0953$+$0755 & J0630$-$2834 & J1136$+$1551 & J0826$+$2637 \\
\hline
2 &	0.335 &	0.031 &	0.221 &	0.359 &	0.309 &	0.332 \\
\hline
3 &	0.350& 	0.037 &	0.249 &	0.382 &	0.332 &	0.360 \\
\hline
4 &	0.361 &	0.041 &	0.276 &	0.402 &	0.351 &	0.385\\
\hline
5 &	0.368 &	0.043 &	0.302 &	0.416 &	0.365 &	0.406\\
\hline
6 &	0.370 &	0.042 &	0.325 &	0.424 &	0.374 &	0.419 \\
\hline
7 &	0.365 &	0.037 &	0.334 &	0.424 &	0.373 &	0.423 \\
\hline
\end{tabular}
\end{centering}
\caption{Table of values of the attenuation parameter $\eta$, defined by eq. (\ref{eqn:etaDef}), for the different sources analyzed. These attenuation effects can be quite large.  HR used only source RX~J1856$-$3754 and J0953+0755. These values are calculated for 100 MeV$\leq E_\gamma \leq $ 400 MeV.  \label{tab:etaPar}} 
%\end{center}
\end{table}

\begin{figure}
\begin{center}
%$\begin{array}{ll}
\subfloat[$n=2$]{\includegraphics[width=0.4\textwidth]{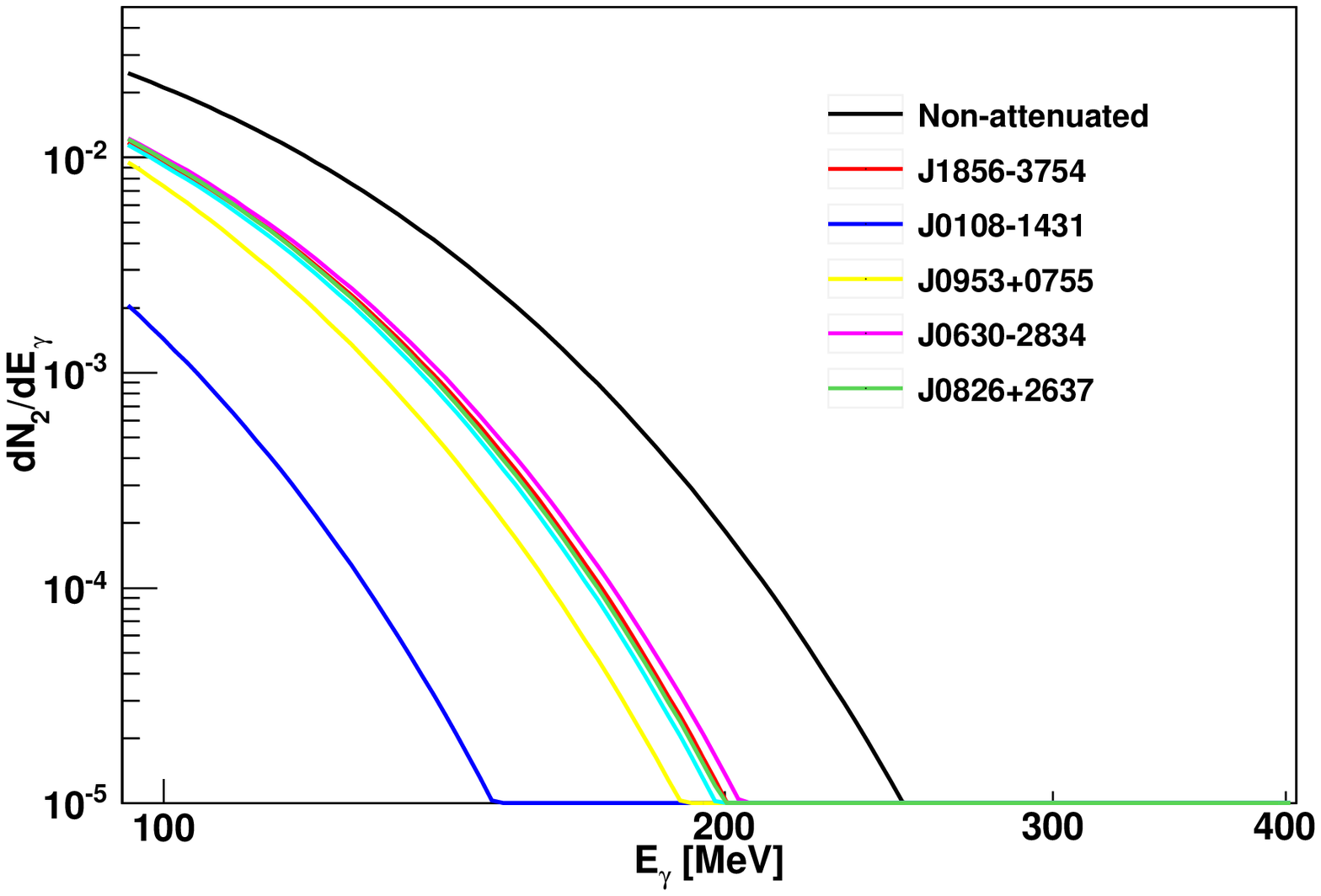}}
\subfloat[$n=5$]{\includegraphics[width=0.4\textwidth]{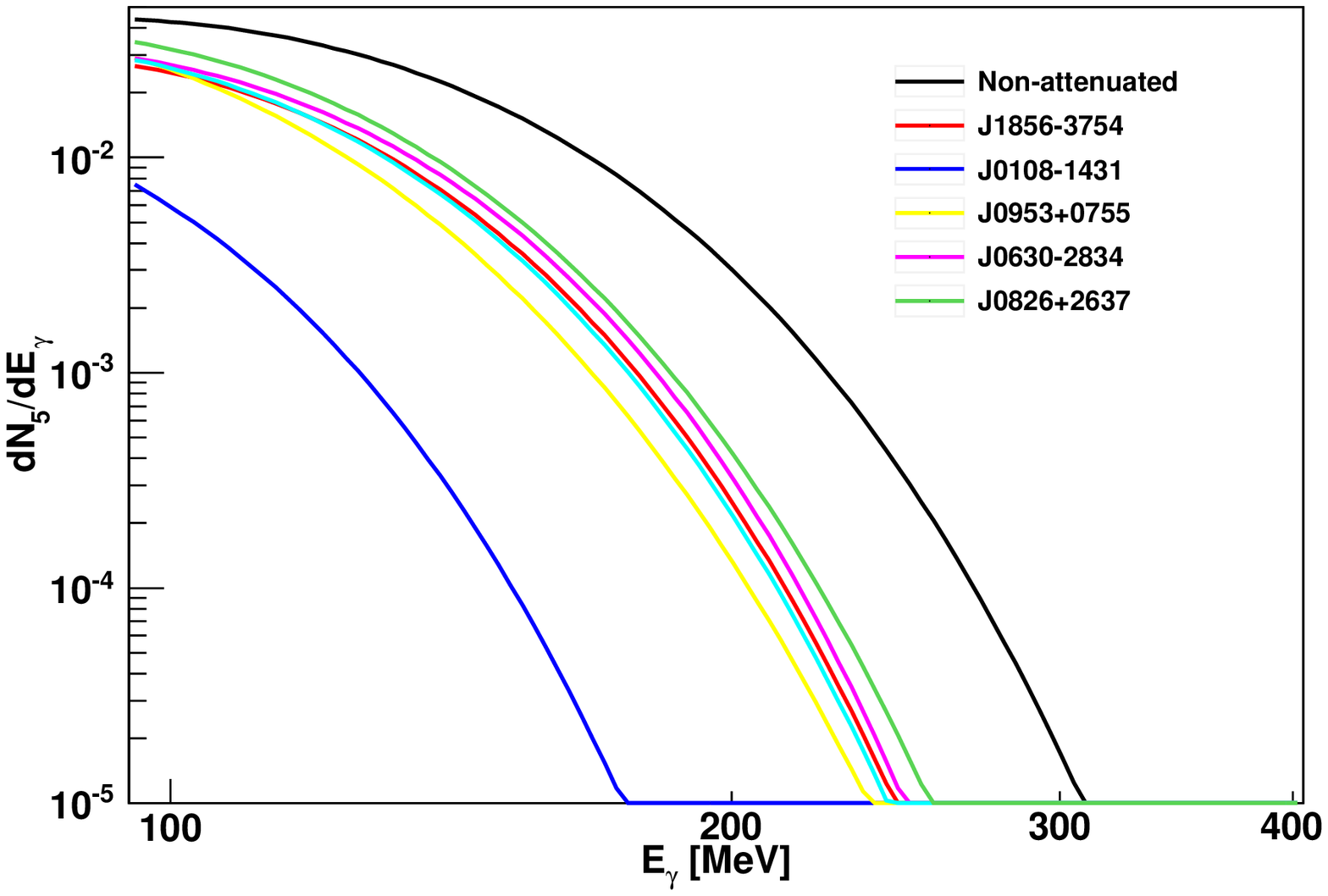}}
% & \includegraphics[width=2.5 in]{compareAtten_testField_script_R20_n5_canvas} 
%\end{array} $
\caption{For $n=2$ and $n=5$, representative distributions of $dN_n/dE_{\gamma}$, according to eq. (\ref{eqn:dNdE}), for the non-attenuated spectrum and all neutron star sources considered, corrected for magnetic pair-production and age attenuation effects.}%Curves are cutoff below $dN_5/dE_\gamma = 10^{-6}$.  
\label{fig:dI_dE}
\end{center}
\end{figure}
%%%%%%%%%%%%%%%%%%%%%%%%%%%%%%%%%%%%%%%%%%%%%%%%%
%%%%%%%%%%%%%%%%%%%%%%%%%%%%%%%%%%%%%%%%%%%%%%%%%
\section{Limits on LED Results}

\subsection{Individual Limits}
With all parameters for the ROI fixed, namely catalog sources and diffuse components, as determined in Section \ref{sec:roi}, upper limits on $R^n$ are determined from spectral fitting based on the method of maximum likelihood.   Fit values are determined by the MINUIT optimizer\cite{MINUITguide}, and one-sided 95\% confidence level upper limits are determined by performing a scan of the log-likelihood function in each ROI\cite{MINUITguide}.  Statistical parameters of the fit are consistent with non-detection of the KK graviton decay signal for all NS considered.%The test statistic (TS) for all neutron star sources is less than 0.001 and is consistent with non-detection, as a significant detection requires a TS$\geq25$\cite{1FGLcatalog}.%, and between -0.01 and 0.01.   % The Planck mass in extra dimensions, $M_{P,n+4}$, is related to the value of $R$, $n$, and the reduced Planck mass $\bar{M}_{P}=2.4\times10^{18}$GeV by:
%The MINOS level is defined as that value of the function -2$\Delta\log{\mathcal{L}(p)}$ such that when reached for a parameter value $p$, MINUIT returns as the MINOS error the value $p-p_{\min}$, where $p_{\min}$ is the fitted value of the parameter.  A MINOS level of 3.84 corresponds to a one-sided 95\% CL upper limit.  Flux upper limits are displayed in Table \ref{tab:fluxLimits}, and the derived limits on $R$ for each $n$ and for each source are displayed in Table \ref{tab:RnLimits}. 
 As a check of this method, we compare upper limits obtained in this manner against upper limits computed using profile likelihood implemented by a different method in the Fermi-LAT \emph{ScienceTools}, and find agreement within 10\%.  Additional systematic checks, due to uncertainties in the parameters of the background sources in the ROI, are also performed to verify the accuracy of the limits, for source RX J1856$-$3754 (the source with the best limits): the agreement in the flux upper limits is found to be 15\% or better.  The flux upper limits for each source and $n$ are displayed in Table \ref{tab:fluxLimits}, and the corresponding limits on the LED size $R$ are shown in Table \ref{tab:RnLimits}. 

\begin{table}
\begin{centering}
\begin{tabular}{|c|c|c|c|c|c|c|}
\hline
n& 	J1856$-$3754& 	J0108$-$1431 &	J0953+0755 &	J1136+1551 &	J0630$-$2834 &	J0826+2637\\
\hline
2 &	3.8 &	4.3 &	5.3 &	4.1 &	5.8 &	6.6\\
\hline
3 &	4.0 &	4.4 &	5.4 &	4.2 &	6.8 &	8.4\\
\hline
4 &	3.7 &	4.2 &	6.2 &	4.4 &	9.4& 	9.9\\
\hline
5 &	4.0 &	4.1 &	6.3 &	4.4 &	11& 	13\\
\hline
6 &	4.1 &	4.0 &	6.7 &	4.2 &	14&	15\\
\hline
7 &	4.2 &	4.0 &	7.8 &	3.5 &	19&	17 \\
\hline
\end{tabular}
\caption{Table of 95\% C.L. flux upper limits ($10^{-9}$ cm$^{-2}$s$^{-1}$) for the sources analyzed.}%, as determined with MINOS errors.}
\label{tab:fluxLimits}
\end{centering}
\end{table}

\begin{table}
\begin{centering}
\begin{tabular}{|l|l|l|l|l|l|l|}
\hline
n & J1856$-$3754 & J0108-14 & J0953+0755 & J1136+1551 & J0630$-$2834 & J0826+2637 \\
\hline
2 &	9.5 &	49  &	22  &	23 &	24 &	29 \\
\hline
3 &	3.9$\times 10^{-2}$ &	0.11 &	6.7$\times 10^{-2}$ &	6.9$\times 10^{-2}$ &	7.4$\times 10^{-2}$ &	8.4$\times 10^{-2}$ \\
\hline
4 &	2.5$\times 10^{-3}$ &	5.4$\times 10^{-3}$  &	3.8$\times 10^{-3}$ &	3.9$\times 10^{-3}$ &	4.3$\times 10^{-3}$ &	4.8$\times 10^{-3}$ \\
\hline
5 &	5.0$\times 10^{-4}$ &	9.1$\times 10^{-4}$ &	7.0$\times 10^{-4}$ &	7.1$\times 10^{-4}$ &	8.1$\times 10^{-4}$ &	8.6$\times 10^{-4}$ \\
\hline
6 &	1.7$\times 10^{-4}$ &	2.8$\times 10^{-4}$ &	2.3$\times 10^{-4}$ &	2.3$\times 10^{-4}$ &	2.7$\times 10^{-4}$ &	2.8$\times 10^{-4}$ \\
\hline
7 &	8.2$\times 10^{-5}$ &	1.3$\times 10^{-4}$ &	1.0$\times 10^{-4}$ &	1.0$\times 10^{-4}$ &	1.2$\times 10^{-4}$ &	1.3$\times 10^{-4}$  \\
\hline
\end{tabular}
\caption{Table of limits on extra dimensions size $R$ (nm) for the sources analyzed.}
\label{tab:RnLimits}
\end{centering}
\end{table}

\subsection{Combined Limits}

\begin{figure}
\begin{centering}
\includegraphics[width=5in]{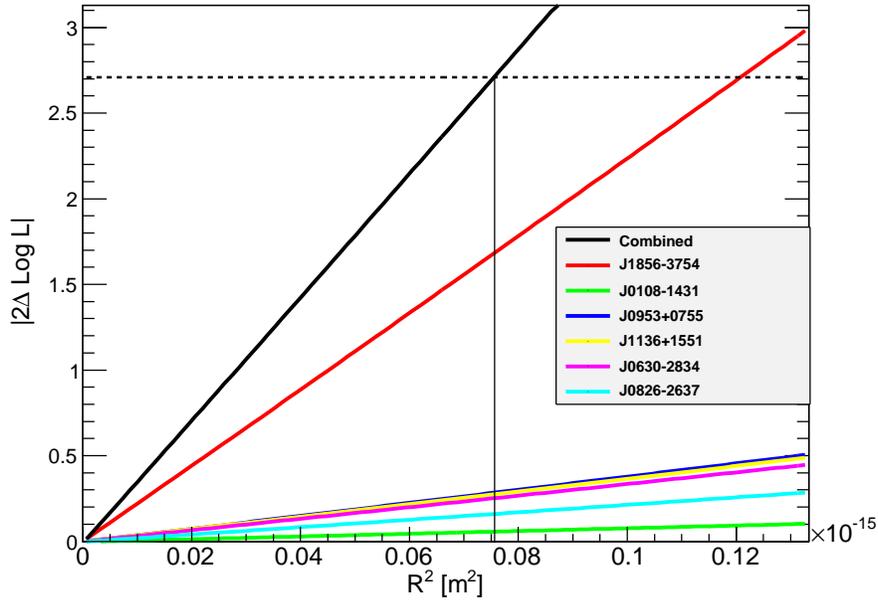}
\caption{\label{fig:Minos_level}  Plot of $|2\Delta\log\mathcal{L}|$ versus parameter value of $R^n$ for $n=2$.  The 95\% confidence level upper limit corresponds to a $y-$axis value of 2.71, shown by a dashed line.  The sum of the curves, solid black, is used to obtain the posterior combined limit at the intersection with 2.71.}
\end{centering}
\end{figure}

We use the following method to combine limits from multiple neutron star sources.
%\subsubsection{Using MINOS error curves}
A scan over the log-likelihood function in each ROI is done with respect to the parameter $R^n$, as shown in Figure \ref{fig:Minos_level}.    A curve of the change in log-likelihood, $|2\Delta\log\mathcal{L}|$, versus parameter value $R^{n}$, is generated for each source.  Then the sum of these curves is taken for all the sources, and the parameter value corresponding to  intersection of that curve with a value of 2.71, corresponding to a one-sided 95\% confidence level, is quoted as the combined limit value.  The results of combining limits on $R$ from this method, as well as results from HR, are presented in Table \ref{tab:Rn}.

\begin{table}
\begin{centering}
\begin{tabular}{|l|l|l|}
\hline
 & $R$ (nm) & $R$ (nm) \\
n & Combined & HR\\
\hline
2 & 8.7 & 51 \\
\hline
3 & 0.037 & 0.11 \\ 
\hline
4 & 2.5$\times 10^{-3}$ & 5.5$\times 10^{-3}$ \\
\hline
5 & 5.0$\times 10^{-4}$ & 9.1$\times 10^{-4}$ \\
\hline
6 & 1.7$\times 10^{-4}$ & 2.8$\times 10^{-4}$ \\
\hline
7 & 8.2$\times 10^{-5}$ & 1.2$\times 10^{-4}$ \\
\hline
\end{tabular}
\caption{95\% CL upper limits on $R$ (nm) for each $n$, compared to HR2003 EGRET-based results\cite{HR2003}.}
\label{tab:Rn}
\end{centering}
\end{table}

\begin{comment}
%\begin{equation} R^{-1} = M_{P,N+4}\left(M_{P,n+4}/\bar{M}_{P}\right)^{2/n} \end{equation}
\begin{table}
\begin{centering}
\begin{tabular}{|l|l|l|l|l|l|l|l|}
\hline
n & Combined (MINOS) & Combined (Likelihood) & HR & CDF & D\O & LEP & ATLAS\footnote{$\Lambda$=1} \\  
\hline
2 & 230 & 97 & 2.09 & 1.40 & 1.60  & \\ 
\hline
3 & 15 &  8.0  & 1.94 & 1.15 & 1.20  & \\
\hline
4 & 2.5 & 1.5 & 1.62 & 1.04 & 0.94 &  \\
\hline
5 & 0.66 & 0.43 & 1.46 & 0.98 & 0.77  & \\ 
\hline
6 & 0.24 & 0.17 & 1.36 & 0.94 & 0.66 & \\
\hline
7 & 0.11 & 0.083 & 1.29 & - & -  & \\
\hline
\end{tabular}

\caption{Comparison of combined limits with previous astrophysical limits and collider limits on $M_D$.}
\label{tab:Planckmass}
\end{centering}
\end{table}
\end{comment}

\subsection{Dependence of LED Limits on Model Parameters}

\begin{comment}
\begin{table}
\begin{centering}
\begin{tabular}{|l|l|l|}
\hline
$n$ & $T = 30 \ \mbox{MeV}$ & $T = 45 \ \mbox{MeV}$ \\
\hline
2 & 6.171E6 & 1.291E8 \\
\hline
3 &5.269E21 & 1.654E23 \\
\hline
4 &4.619E36 & 2.175E38 \\
\hline
5 &.259E51 & 3.008E53 \\
\hline
6 & 4.151E66 & 4.397E68 \\
\hline
7 & 4.273E81 & 6.790E83 \\
\hline
\end{tabular}
\caption{Table of values of $k_n$, evaluated at two different values of $T$.}
\label{tab:kn_Tdep}
\end{centering}
\end{table}
\end{comment}

%\begin{comment}
\begin{table}
\begin{centering}
\begin{tabular}{|l|l|l|}
\hline
n & $T = 30 \ \mbox{MeV}$ & $T = 45 \ \mbox{MeV}$ \\
\hline
2 & $9.5$ & $1.2$ \\
\hline
3 & $3.9\times 10^{-2}$ & $9.3\times 10^{-3}$ \\
\hline
4 & $2.5\times 10^{-3}$ & $8.5\times 10^{-4}$ \\
\hline
5 & $5.0\times 10^{-4}$ & $2.1\times 10^{-4}$ \\
\hline
6 & $1.7\times 10^{-4}$ & $8.1\times 10^{-5}$ \\
\hline
7 & $8.2\times 10^{-5}$ & $4.1\times 10^{-5}$ \\
\hline
\end{tabular}
\caption{A comparison of upper limits on $R$ (nm), evaluated for different values of $T$, for source RX~J1856$-$3754.}
\label{tab:R_Tdep}
%\end{tabular}
\end{centering}
\end{table}
%\end{comment}

%Dependence of the limits on model parameters, namely $T$, $f_{KK},$ and $\delta t_{NS}$, have been evaluated.  We have determined that the bounds on extra dimensions are quite sensitive to change in $T$.  Limits evaluated for source RX J1856$-$3754 for $T=30$ MeV and a higher value $T=45$ MeV, are compared in Table \ref{tab:R_Tdep}. The limits on LED are a strong function of temperature.  The dependence enters through two effects: changing the constant $k_n$ and changing the distribution of gamma-ray energies.  The limits are affected, since $k_n \sim T^{-n-5.5}$; in other words, by modifying $k_n$, the bounds on LED size improve as:
Dependence of the limits on model parameters, namely $T$, $f_{KK},$ and $\Delta t_{NS}$, have been evaluated.  We have determined that the bounds on extra dimensions are quite sensitive to changes in $T$.  Limits evaluated for source RX J1856$-$3754 for $T=30$ MeV and a higher value $T=45$ MeV, are compared in Table \ref{tab:R_Tdep}. The limits on LED are a strong function of temperature.  The dependence enters through two effects: changing the constant $k_n$ and changing the distribution of gamma-ray energies.  The limits are affected, since $k_n \sim T^{-n-5.5}$; in other words, by modifying $k_n$, the bounds on LED size improve as:
\begin{equation} R \sim \left(\frac{T}{30 \ \mbox{MeV}}\right)^{-1-5.5/n}. \end{equation}  

%Values of $k_n$, evaluated at $T = 30, 45$ MeV, are compared in Table \ref{tab:kn_Tdep}. 
 In addition,  for higher temperatures, the distribution of energies is shifted to higher gamma-ray energies.  Quantitatively, this increases the integral of the distribution function above 100 MeV, $\int_{100 \mbox{MeV}}^{\infty} dN/dE \ dE$, which tends to improve the bounds.  %Sample curves are shown in Figure \ref{fig:n5_Tdep}. 
 Limits placed on $R$ from source RX J1856$-$3754 may vary by an order of magnitude, as shown in Table \ref{tab:R_Tdep}.   We do not consider lower values of $T$, since according to \cite{raffeltBook}, $T=30$ MeV is a conservative lower limit on the SN core temperature.  

%By varying the timescale of core collapse, $\Delta t_{NS}$, the limits on $R$ vary as $\left(\Delta t_{NS}\right)^{-1/n}$.   Estimates for this parameter vary from 5 s to 20 s\cite{RosswogBruggen}.  Varying the fraction of energy lost into gravitons channel, $f_{KK}$, the limits on $R$ vary as $f_{KK}^{-1/n}$, according to HR.  $f_{KK}\simeq 0.01$ is determined to be consistent with SN1987A \emph{and} diffuse measurements according to EGRET\cite{egretDiffuse,HR_SN_diffuse}. Thus, we see that the limits depend only weakly on variations of these two parameters.

By varying the timescale of core collapse, $\Delta t_{NS}$, the limits on $R$ vary as $\left(\Delta t_{NS}\right)^{-1/n}$.   Estimates for this parameter vary from 5 s to 20 s\cite{RosswogBruggen}, while we use the value of 7.5 s from HR.  Thus, we see that the limits depend only weakly on variations of $\Delta t_{NS}$ and $f_{KK}$.

\begin{table}
\begin{centering}
\begin{tabular}{|l|l|l|}
\hline
$n$ & $f_{KK}/10^{-3}$ & $R$ (nm) \\
\hline
2  &	6.3 & 	9.5 \\
\hline
3 &     8.7  &	0.035 \\
\hline
4 &	7.4 &	2.5 $\times 10^{-3}$ \\
\hline
5 &	5.1 &	5.3 $\times 10^{-4} $\\
\hline
6 &	9.1 &   1.7 $\times 10^{-4}$\\
\hline
7 &	9.0 &	8.0 $\times 10^{-5}$ \\
\hline
\end{tabular}
\caption{\label{tab:fKKprior1}Table of $f_{KK}$ values and combined upper limits on  $R$ (nm) for each value of $n$, assuming a Gaussian prior on $f_{KK}$ (with mean 0.0075 and sigma 0.00144), as discussed in Sec. \ref{sec:fKKlimit}.}
\end{centering}
\end{table}

\subsection{Effect of Uncertainties on $f_{KK}$ on the Limits}\label{sec:fKKlimit}
Varying the fraction of energy lost into the graviton channel, $f_{KK}$, the limits on $R$ vary as $f_{KK}^{-1/n}$.  HR assumed $f_{KK}\simeq 0.01$, as consistent with diffuse gamma-ray measurements according to EGRET\cite{egretDiffuse,HR_SN_diffuse}. However, a more accurate treatment from EGRET low energy diffuse measurements constrains $f_{KK}$ such that $0.005<f_{KK}<0.01$.  To take this range of values for $f_{KK}$ into account when computing limits, we perform an analysis allowing for a Gaussian prior on the $f_{KK}$ parameter, with a mean of 0.0075 and a sigma of 0.00144 (as obtained from the variance for a uniform PDF for $f_{KK}$ between 0.005 and 0.01).  We constrain this parameter to be the same across the 6 ROIs, for each value of $n$.  This is possible within the framework of the Fermi-LAT \emph{ScienceTools};  a similar technique was used to constrain dark matter signals from a combined analysis of Milky Way satellites with the Fermi-LAT\cite{dwarfStack}.  Limits obtained in this manner are shown in Table \ref{tab:fKKprior1}.

%Furthermore, HR demonstrated that the $f_{KK}$ does not change very much for $T \geq$ 30 MeV.  

\section{Discussion and Conclusions\label{sec:int_and_concl}}
If $M_D$ is at a TeV, then for $n<4$, the results presented here imply that the compactification topology is more complicated than a torus, i.e., all LED having the same size. For flat LED of the same size, the lower limits on $M_D$ results are consistent with $n\geq4$. The constraints on LED based on neutron star gamma ray emission yield improvements over previously reported neutron star limits, based on gamma-ray measurements and combination of individual sources, as shown in Table \ref{tab:Planckmass}.  In addition, the results for the $n$-dimensional Planck mass are much better than collider limits from LEP and Tevatron for $n<4$, and are comparable or slightly better for $n=4$.% as shown in Table \ref{tab:Planckmass}.  

\begin{table}
\begin{centering}
\begin{tabular}{|l|l|l|l|l|l|l|l|l|}
\hline
n & Combined & CDF & D\O & LEP & ATLAS & CMS \\
\hline
2 & 230 & 2.09 & 1.40 & 1.60  & 1.5 & 3.2  \\
\hline
3 & 16 &  1.94 & 1.15 & 1.20  & 1.1 & 3.3 \\
\hline
4 & 2.5 & 1.62 & 1.04 & 0.94 &  1.8 & 3.4  \\
\hline
5 & 0.67 & 1.46 & 0.98 & 0.77  & 2.0 & 3.4\\
\hline
6 & 0.25 & 1.36 & 0.94 & 0.66 & 2.0  & 3.4\\
\hline
7 & 0.11 & 1.29 & - & -  & - & - \\
\hline
\end{tabular}
\caption{Comparison of 95\% CL lower limits on $M_D$ (TeV) with previous astrophysical limits and  collider limits.  \emph{Combined} limits are obtained in this paper.  Collider limits are taken from references \cite{LHC_LED,LEP_LED,D0_LED,CDF_LED}.  ATLAS and CMS results are quoted where $\Lambda/M_D=1$.   ATLAS results are quoted with 3.1 pb$^{-1}$ of data; CMS results are quoted with  36 pb$^{-1}$ of data.\label{tab:LLcompare}}
\label{tab:Planckmass}
\end{centering}
\end{table}

These limits may prove useful, especially for $n=4$ case (where the limits are comparable to collider results), in the context of constraining phase space in searches for extra dimensions underway at the LHC.  These results are also more stringent than those reported by short distance gravity experiments probing for deviations from the inverse square law.  The most sensitive such experiment to KK graviton emission presented a result of 37 $\mu$m for $n=2$ at 95\% C.L.\cite{eotwash2006}; this is several orders of magnitude larger than the combined result reported here, of 8.7 nm. %  In addition, constraints on LED may be able to be translated into certain scenarios of warped extra dimensions, such as weakly warped extra dimensions for $n=2$.  

Cass\'e \emph{et al.} obtain upper limits significantly better than ours~\cite{Casse2004} (a factor $\sim 20$ for $n=2$, though decreasing approximately as $1/n^2$ with increasing $n$), summing the contribution of all the expected NS in the Galactic bulge and comparing to the EGRET data. But it is should be noted that they do not account for age nor magnetic field attenuation, while the present analysis shows that both impact the photon distribution in a significant way. As these effects are not taken into account in HR, which the Cass\'e \emph{et al.} paper is based on, their upper limits are necessarily underestimated. Furthermore, it should be emphasized that the 6 neutron stars analyzed here are chosen at high latitude to avoid the large systematic uncertainties involved in modeling the diffuse Galactic background. These are even larger in the low energy range that we are interested in here. An analysis of the bulge with the current Fermi-LAT instrument response functions and the current Fermi-LAT diffuse models could nominally improve the flux upper limits, but at the cost of a much less robust analysis, the systematics of which are difficult to evaluate.   Within the Fermi-LAT collaboration, a better inner galaxy model is in process, which is necessary before approaching a Galactic Bulge analysis.  A Large Extra Dimensions analysis of the Galactic Bulge will be the subject of future studies.

\begin{comment}
It is not practical to perform a GB analysis in the method of Cass\'e et al., taking into account the magnetic field attenuation (and for that matter, the effects of neutron star ages).   We do not have a distribution function of magnetic field strength over the appropriate sample of NSs.   As we have shown here, the attenuation effects due to the magnetic field is not small.  NS ages have a similar consideration.  These effects are not taken into account in HR, and thus there is now a proven underestimation of the upper limits in HR, and consequently in Cass\'e et al.

It is important to note that in this work, the 6 neutron stars are chosen at high latitude to avoid the large systematics involved in modeling the diffuse Galactic background. This is even more true at the low gamma-ray energies of this analysis, so that an analysis of the bulge with the current Fermi-LAT instrument response functions and the current Fermi-LAT diffuse models could nominally improve the flux upper limits, but at the cost of a much less robust analysis. Within the Fermi-LAT collaboration, a better inner galaxy model is in process, which is necessary before approaching a Galactic Bulge analysis.  An analysis of the bulge will be the subject of future studies.

\end{comment}

\appendix

\section{Appendix: Sampling Decay Vertices from Graviton Trajectories}\label{sec:appendixA}

In our model, due to the $G_{KK}$ emission radially outward during the SN core collapse, the $G_{KK}$ are not given any initial angular momentum; thus we assume that the $G_{KK}$ oscillate on radial paths (completely eccentric orbits) through the center of the neutron star.  In spherical coordinates $(r,\theta,\phi)$, this is equivalent to the following constraints: $\dot{\theta}=\dot{\phi}=0$.  The orbital radial distribution, $P(r;\mu)$, is defined outside the neutron star by the radial Kepler equation, in which time is given as a function of the radial coordinate $r$\cite{goldstein}\footnote{The radial Kepler equation is not manifestly periodic.  However, the full orbit cycle includes an interval over $0<r<r_{\max}$, or a quarter of a cycle.  Due to the symmetry of the orbit, our treatment of sampling the decay vertex from $t=0$ to $t(r_{\max})$ is sufficient to obtain the full distribution of $r_0$.}:
\begin{equation} t(r) = t_k\left(\arcsin\left(\sqrt{kr}\right)-\sqrt{kr(1-kr)}+c_1\right), \end{equation}

where:
\begin{equation} k = r_{\max}^{-1} = \frac{1+1.5|U_{NS}|-\gamma}{|U_{NS}|}\ ,\label{eqn:rmax}\end{equation}
and:
\begin{equation} t_k = \frac{R_{NS}}{\beta c k^2}\sqrt{\frac{k(1-k)}{1-\left|U_{NS}\right|/\beta^2}}. \end{equation}
The solution inside the NS ($r<1$) is defined as:
\begin{equation} r(t) = \frac{\beta}{\sqrt{\left|U_{NS}\right|}} \sin(\Omega t)\ , \end{equation}

\noindent where the parameter $\Omega = \sqrt{|U_{NS}|}c/R_{NS} = 9.13\times10^3$ s$^{-1}$.  

\begin{figure}
\begin{centering}
\includegraphics[width=0.6\textwidth]{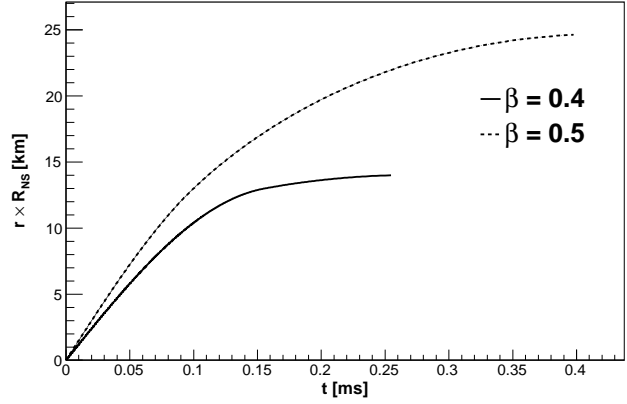}
%\end{tabular}
\caption{Trajectories for $\beta = 0.4$ and $\beta=0.5$. Notice that for larger $\beta$, the KK graviton may achieve a farther maximum distance, $r_{\max} \times R_{NS}$.  This figure shows one quarter of the orbit cycle.}
 \label{fig:TrajectoriesBeta1}
\end{centering}
\end{figure}

%Due to different dependence on $r$ inside and outside the neutron star, due to the gravitational potential, the trajectory as a function of time is different inside and outside.  We determine radial distributions, $P(r;\mu)$, for 50 values of $\mu$, linearly-spaced between $\mu_{\min}$ and $\mu_{\max}$.  
Given that $t=0$ is the time when the $G_{KK}$ is created at $r=0$, the radial distributions are determined by sampling time uniformly between the $t=0$ and $t_{\max}$. $t_{\max}$ is given by the time to achieve the maximum distance, $r_{\max} = 1/k$, as in eq. (\ref{eqn:rmax}).  $c_1$ is determined from boundary conditions of position and velocity at the surface of the neutron star by solving the full equation of motion inside and outside the neutron star.  The trajectories for a couple of values of $\beta$ are plotted in Figure \ref{fig:TrajectoriesBeta1}, while the radial distributions for representative values of $\mu$, $P(r;\mu)$, are shown in Figure \ref{fig:RadialDistributions}.  %The radial distribution for $n=2$, averaged over all $\mu$, is displayed in Figure \ref{fig:radialProfile}.  

\section{Appendix: Relativistic Decay Kinematics of KK Gravitons}\label{sec:appendixB}
\begin{figure}
\begin{centering}
\includegraphics[trim=2in 3in 2in 3in, width=0.7\textwidth]{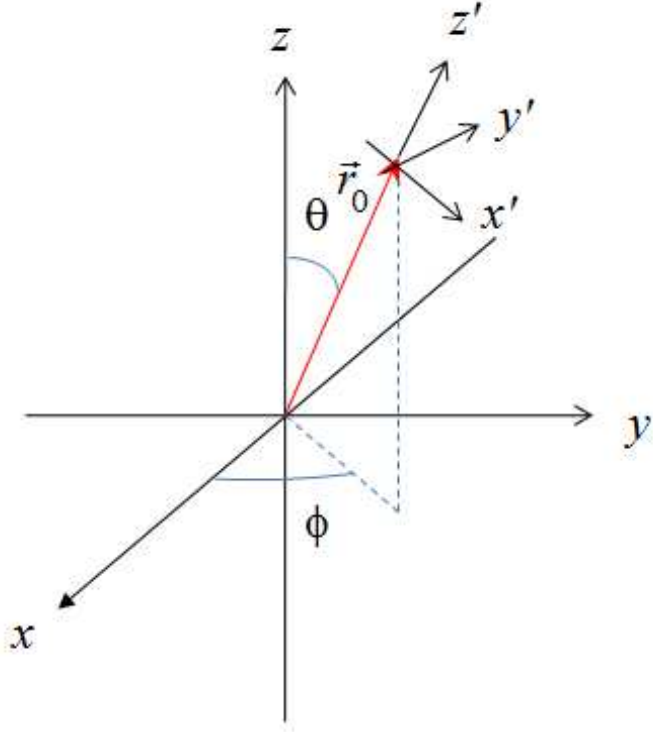}
\caption{The coordinate system, as described in Appendix \ref{sec:appendixB}.}
\label{fig:coordSys}
\end{centering}
\end{figure}

The energy is given by:
	\begin{equation} E_\gamma = \frac{1}{2}\gamma^\prime m\left(1+\beta^\prime\cos\theta^*\right), \end{equation}
while the components of the photon momentum, 
\begin{equation} \vec{p}_\gamma^\prime =  p_{x^\prime,\gamma}\hat{x}^\prime \ + \  p_{y^\prime,\gamma}\hat{y}^\prime \ + \ p_{z^\prime,\gamma}\hat{z}^\prime,\end{equation}
 in the neutron star frame relative to the direction of the $G_{KK}$ are given by:
\begin{align} p_{x^\prime,\gamma} &= \frac{1}{2}m\sin\theta^*\cos\phi^* \\
				p_{y^\prime,\gamma} &=  \frac{1}{2}m\sin\theta^*\sin\phi^* \\
				p_{z^\prime,\gamma} &= \frac{1}{2}\gamma^\prime m(\beta^\prime+\cos\theta^*).  \end{align}
The $z^\prime$ axis is defined by: $\hat{z}^\prime=\hat{r}_0$.  $\theta^*$ is the polar angle between the direction of the $G_{KK}$ in the lab frame ($z^\prime$) and the decay photon in the rest frame of the $G_{KK}$, and $\phi^*$ is the angle in the ($x^\prime-y^\prime$)plane. In this frame, $\hat{x}^\prime$ is taken as perpendicular to $\hat{z}^\prime$ in the $z-z^\prime$ plane and in the direction of increasing $\theta$, and $\hat{y}^\prime = \hat{z}^\prime \times \hat{x}^\prime $.  The coordinate systems used are depicted in Fig. \ref{fig:coordSys}.  %This relation is handled by the delta-function in the integrand of eqn. (\ref{eqn:MCintegration}).  
We sample a $\cos\theta^*$ value uniformly over the interval [-1,1], and $\phi^*$ uniformly over the interval [$-\pi,\pi$], given isotropic emission of photons in the rest frame of the gravitons. % We obtain momentum components of the decay gamma-rays in the rest frame of the graviton, and
Subsequently, we obtain momentum components of the gamma ray in the frame of the neutron star, $\vec{p}_\gamma$, 
%btain the momentum components in the frame of the neutron star, it is
by rotating back into the frame of the neutron star, using the direction of the momentum vector, $\vec{p}_{KK}$, as defined by step (5).  This is needed for the next step.  %This is a rotation in passive sense, in that the reference coordinate system is changing, not the direction of the vector. %This process is implemented with the TVector3::RotateUz function in ROOT.   % with the angles ($\theta,\phi$).   

\section{Appendix: Determining Photon Pair Production Optical Depths} \label{sec:appendixC}

Approximations for the pair-production attenuation are according to the treatment in Refs. \cite{Erber1966,landauLifshitzBook}.  The attenuation coefficient depends on the parameter: 

\begin{equation} \chi(E_\gamma,\vec{p}_\gamma,\vec{r}) = \frac{E_{\gamma}}{m_e c^2} \frac{B_{\perp}(\vec{r},\vec{p}_\gamma)}{B_{\tmop{cr}}},\end{equation} 
where the critical field is given by 

\begin{equation} B_{\tmop{cr}} = \frac{m^2_{e^{}} c^3}{e \hbar} = 4.414 \times 10^{13} \ \mbox{G}, \end{equation}

\begin{figure}
\begin{centering}
\includegraphics[width=0.7\textwidth]{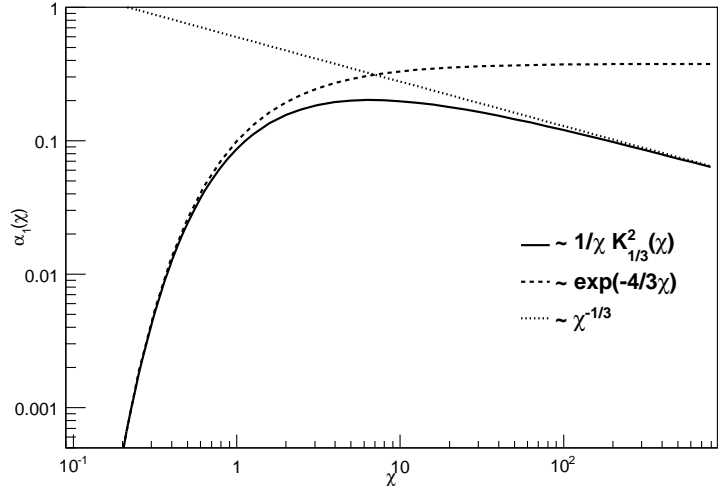}
\caption{\label{fig:attenCurve} Plot of the reduced attenuation coefficient $\alpha_1(\chi)$ and limiting asymptotic expressions corresponding to eq. (\ref{eqn:attenEqn}). } 
\end{centering}
\end{figure}

 $B_{\perp}$ is the magnetic field component of the neutron star perpendicular to the photon's momentum vector $\vec{p}_\gamma$, and $E_{\gamma}$ is the photon energy. \ For the magnetic field of the neutron star, we assume a static dipole field, 

\begin{equation} \vec{B}(\vec{r}) = \frac{3(\vec{m}\cdot\hat{r})\hat{r}-\vec{m}}{r^3}\end{equation}

with dipole moment $\vec{m}=\frac{1}{2} B_{surf}R_{NS}^3\hat{z}$.  The attenuation coefficient, $\alpha$, is given by: \begin{equation}
  \alpha (\chi(E_\gamma,\vec{p_\gamma},\vec{r})) = \frac{\alpha_{\tmop{fs}}}{\lame}
  \frac{B_{\perp}(\vec{r},\vec{p}_\gamma)}{B_{\tmop{cr}}} \alpha_1(\chi),
\end{equation}

where the reduced attenuation coefficient, $\alpha_1(\chi)$, is expressible as a function of $\chi$ in terms of a modified Bessel function of the second kind, with asymptotic limiting expressions for small and large values of $\chi$ (as plotted in Figure \ref{fig:attenCurve}):

\begin{equation}
\alpha_1(\chi) = 0.16 \frac{1}{\chi} K_{1 / 3}^2 \left( \frac{2}{3
  \chi} \right) = \begin{array}{c}
    
  \end{array} \left\{ \begin{array}{l}
    0.377 e^{- \frac{4}{3 \chi}}, \chi \leq 0.1\\
    \\
    0.597 \chi^{- 1 / 3}_{}, \chi \geq 100
  \end{array} \right.
\label{eqn:attenEqn}
\end{equation}
 In eq. (\ref{eqn:attenEqn}), $\lame=3.861\times10^{-11}$ cm is the reduced electron Compton wavelength and $\alpha_{\tmop{fs}}$ is the fine structure constant.  The asymptotic expressions are used in the Monte Carlo simulation, where appropriate, in order to save computer time.

The optical depth $\tau_{pp}$ is calculated by path integrating the attenuation coefficient along the direction of the photon, from the point of decay, $\vec{r}_0\equiv<x_0,y_0,z_0>$, out to where $r_{km}=7 R_{NS}$ (where the field has attenuated to 0.3\% of the surface field strength), according to 
\begin{equation}
\begin{split} \tau_{pp} &= \int_{\tmop{path}}\alpha \ \dd s \\
			&= \int_{0}^{s_{\max}} \alpha\left(\chi\left(E_\gamma,\vec{p}_\gamma,\vec{r}_0+\hat{p}_\gamma s\right)\right) \dd s \\
\end{split}
\end{equation}

In the preceding equation, $s_{\max}$, given by:

\begin{equation} s_{\max}=-x_0 p_1-y_0 p_2 -z_0 p_3 + \sqrt{(x_0p_1+y_0p_2+z_0p_3)^2+(7R_{NS})^2-x_0^2-y_0^2-z_0^2}, \end{equation}

\noindent refers to the path length where the photon with direction unit vector 
*
\begin{equation} \hat{p}_\gamma=\frac{\vec{p}_\gamma}{\left|\vec{p}_\gamma\right|}=p_1 \hat{x} + p_2 \hat{y} + p_3 \hat{z} \end{equation} 
\noindent is  considered to have escaped the magnetosphere.

\begin{comment}
The constant $t_k$ is determined analytically, and $c_1$ is determined from boundary conditions of position and velocity at the surface of the neutron star.  The trajectories for different values of the initial value of $\beta$ are plotted in Figure \ref{fig:TrajectoriesBeta1}, while the radial probability distributions for particular values of $\mu$ are shown in Figure \ref{fig:RadialDistributions}.  The radial distribution for $n=2$, averaged over all $\mu$, is shown in Figure \ref{fig:radialProfile}
\end{comment}

\begin{acknowledgments}

The \textit{Fermi}-LAT Collaboration acknowledges generous ongoing support
from a number of agencies and institutes that have supported both the
development and the operation of the LAT as well as scientific data analysis.
These include the National Aeronautics and Space Administration and the
Department of Energy in the United States, the Commissariat \`a l'Energie Atomique
and the Centre National de la Recherche Scientifique / Institut National de Physique
Nucl\'eaire et de Physique des Particules in France, the Agenzia Spaziale Italiana
and the Istituto Nazionale di Fisica Nucleare in Italy, the Ministry of Education,
Culture, Sports, Science and Technology (MEXT), High Energy Accelerator Research
Organization (KEK) and Japan Aerospace Exploration Agency (JAXA) in Japan, and
the K.~A.~Wallenberg Foundation, the Swedish Research Council and the
Swedish National Space Board in Sweden.

Additional support for science analysis during the operations phase is gratefully
acknowledged from the Istituto Nazionale di Astrofisica in Italy and the Centre National d'\'Etudes Spatiales in France.

%This work was supported by the US Department of Energy under SLAC National Accelerator Laboratory contract DE-AC03-765F00515.

\end{acknowledgments}
\bibliographystyle{JHEP}
\bibliography{ref_JCAP}

\end{document}